\documentclass[12pt,preprint]{aastex}
\def\wise{{\it WISE }}
\def\spitz{{\it Spitzer }}
\def\mw{W}

\begin{document}

\shortauthors{Esplin, Luhman \& Mamajek}
\shorttitle{\wise Survey of Taurus}

\title{A \wise Survey of Circumstellar Disks in Taurus\altaffilmark{1} }

\author{
T. L. Esplin\altaffilmark{2}, 
K. L. Luhman\altaffilmark{2,3} and 
E. E. Mamajek\altaffilmark{4} 
}

\altaffiltext{1}
{Based on data from the {\it Wide-field Infrared Survey Explorer}, the {\it Spitzer Space Telescope}, the Two Micron All-Sky Survey,
the NASA Infrared Telescope Facility, the Hobby-Eberly Telescope,
and the Digitized Sky Survey.}
\altaffiltext{2}{Department of Astronomy and Astrophysics, The Pennsylvania
State University, University Park, PA 16802; taran.esplin@psu.edu.}
\altaffiltext{3}{Center for Exoplanets and Habitable Worlds,
The Pennsylvania State University, University Park, PA 16802.}
\altaffiltext{4}{Department of Physics and Astronomy, The University of Rochester, Rochester, NY 14627, USA}

\begin{abstract}
We have compiled photometry at 3.4, 4.6, 12 and 22 $\mu$m  
from the all-sky survey performed by the {\it Wide-field Infrared Survey Explorer} ({\it WISE})  
for all known members of the 
Taurus complex of dark clouds.
Using these data and photometry from the {\it Spitzer Space Telescope},
we have identified members with infrared excess emission from circumstellar 
disks and have estimated the evolutionary stages of the detected disks,
which include 31 new full disks and 16 new candidate transitional, evolved, 
evolved transitional, and debris disks. 
We have also used the \wise All-Sky Source Catalog to search for new
disk-bearing members of Taurus based on their red infrared colors.
Through optical and near-infrared spectroscopy,
we have confirmed 26 new members with spectral types of M1 -- M7. 
The census of disk-bearing stars in Taurus should now be largely complete
for spectral types earlier than $\sim$M8 ($M \gtrsim 0.03$~$M_\odot$).

\end{abstract}


\section{Introduction}

Complete samples of circumstellar disks in star-forming regions
and accurate classifications of those disks represent a foundation for
studies of star and planet formation.
Both the identification of circumstellar disks and their classification
are most easily performed using mid-infrared (IR) continuum emission from
warm circumstellar dust.
Because the atmosphere is bright and strongly absorbing at mid-IR wavelengths,
space-based telescopes have provided the most sensitive data of this kind.
The all-sky mid-IR images from the {\it Infrared Astronomical Satellite}
\citetext{IRAS; \citealt{neu84}} enabled the first detections of circumstellar
disks in star-forming regions. The most thorough census of disks was
performed in Taurus because it is nearby ($d=140$~pc; \citealt{wic98,loi05,tor07,tor09})
and has a low enough
stellar density that its members could be resolved by {\it IRAS} \citep{ken95}.
Subsequent mid-IR telescopes, such as the {\it Infrared Space Observatory}
 \citep{kes96} and the {\it Spitzer Space Telescope}
 \citep{wer04}, have offered progressively better sensitivity and resolution,
enabling detections of disks at fainter levels and in more crowded fields.
Due to its modest field of view (5$\arcmin$), 
\spitz primarily observed more compact star-forming regions 
(\citealt{lad06}; \citealt{sic06}; \citealt{dah07};
\citealt{her07}; \citealt{luh08}; \citealt{gut09}),
although it was able to map a significant fraction of a few widely distributed
populations such as Taurus 
(\citealt{luh10tau}; \citealt{reb10}).

The latest mid-IR satellite, the {\it Wide-field Infrared Survey Explorer} 
({\it WISE}; \citealt{wri10}),
has lower spatial resolution than \spitz
but it covered the entire sky.
As a result, \wise provides mid-IR photometry  
for the portions of large star-forming regions and 
associations that were not imaged by {\it Spitzer}.
For instance,
\wise data have been used 
to search for new disk-bearing stars in Taurus \citep{reb11}
and
to classify disks among the known members of 
Upper Sco \citep{luh12}.
Because of the importance of Taurus in studies of disks
and because proper use of \wise data requires great care,
we have performed our own search for new members with disks in Taurus,
and have classified all the disks that we have detected around the known members. 
In this paper
we begin by compiling photometry from 3 to 24 $\mu m$ 
from both {\it WISE} and \spitz for all known members of Taurus
(Section \ref{sec:photo}).  With these data, we then measure 
mid-IR excesses (Section \ref{sec:excess})
and classify the evolutionary stages of the detected disks
(Section \ref{sec:class}).  Finally, we
use \wise photometry in conjunction with proper motions, color-magnitude
diagrams, and follow up spectroscopy to identify new members of Taurus
(Section \ref{sec:search}).

\section{Photometric Data}
\label{sec:photo}

\subsection{Known Members of Taurus}
\label{sec:members}
\begin{figure}[h]
	\centering
	\includegraphics[scale=.7]{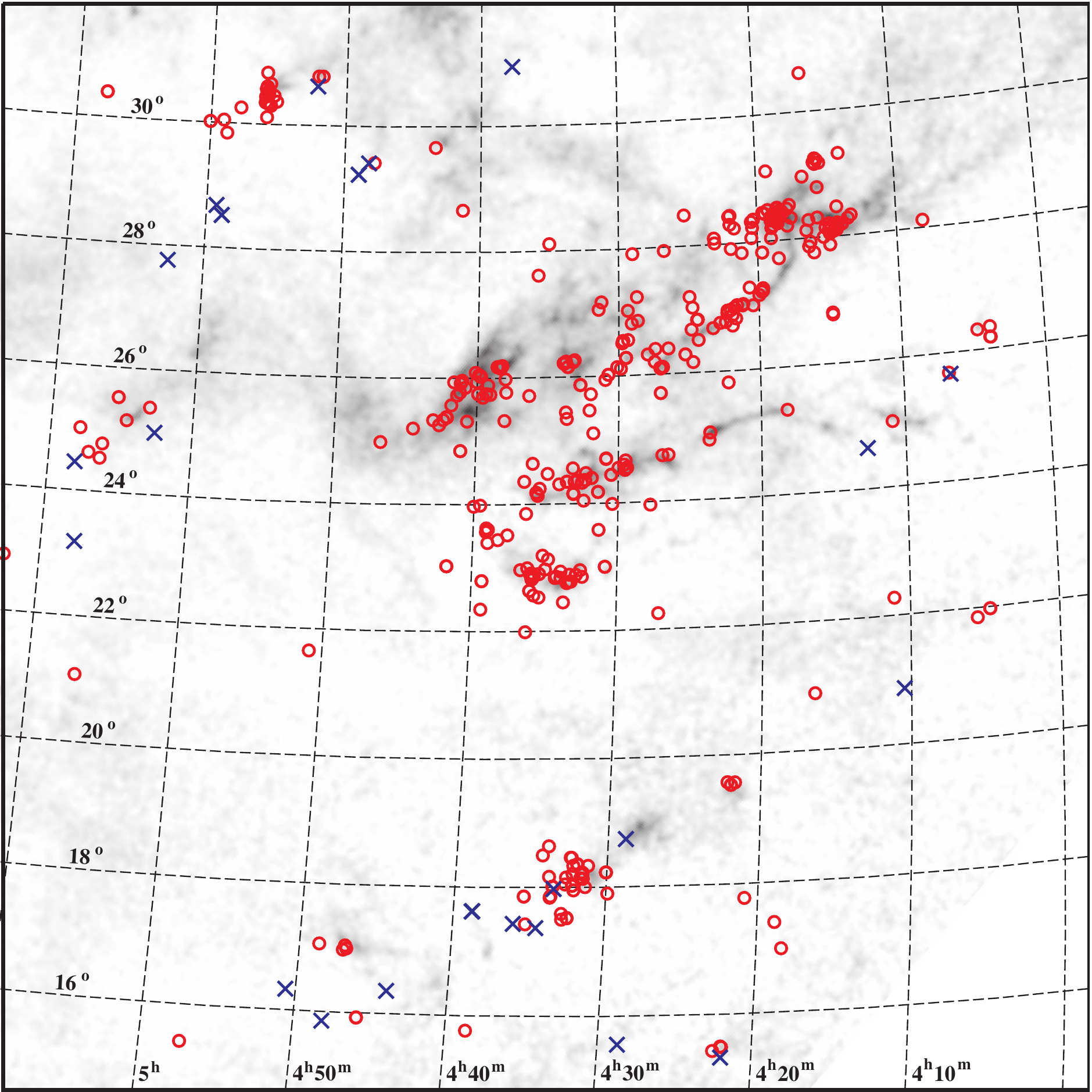}
\caption{Spatial distribution of previously known members of the Taurus star-forming region (circles) and new members from this work (crosses). The dark clouds in Taurus are displayed with a map of extinction (gray scale;  \citealt{Dob05}).}
\label{fig:map}
\end{figure}

To construct a census of the circumstellar disks in Taurus,
we begin by compiling a list of all known members of the region.
We adopt the 352 members from \cite{luh10tau},
four additional stars that have good evidence of membership from previous studies
(Section \ref{sec:newmembers}),
32 members found in a subsequent survey by Luhman (in preparation; 33 if GZ Tau A and B 
are counted separately),
and 25 new members 
(26 if BS Tau A and B are counted separately)
that we have confirmed with spectroscopy (Section \ref{sec:spectra}).
We also adopt as a member HD 285957, which has a proper motion consistent 
with that of the Taurus subgroup L1551 \citep{zach12,luh09} and exhibits
evidence of youth in the form of Li absorption \citep{wic00,ses08}.
Known multiple systems are treated as single sources in our census
unless they are resolved by either \spitz or the 
Two Micron All-Sky Survey \citep[2MASS;][]{skr06}. 
The resulting catalog contains 414 sources and is presented in
Table \ref{tab:all}. The spatial distribution of these members is
illustrated in a map of the Taurus dark clouds in Figure \ref{fig:map}.

\subsection{\spitz Photometry}

We make use of mid-IR photometry for members of Taurus 
measured with {\it Spitzer's} 
Infrared Array Camera (IRAC; \citealt{faz04}) and the 
Multiband Imaging Photometer for \spitz (MIPS; \citealt{rie04}).
We consider the four bands of IRAC (3.6, 4.5, 5.8, and 8.0 $\mu$m) 
  and the 24 $\mu$m band of MIPS, 
which are denoted as [3.6], [4.5], [5.8], [8.0], and [24], respectively.
IRAC produced images with a field of view of $5\farcm2 \times 5\farcm2$
and FWHM of $1\farcs6-1\farcs9$ for [3.6] to [8.0].  
MIPS had a field of view of $5\farcm4 \times 5\farcm4$ and a FWHM of $5\farcs9$
for [24].

Photometry from most \spitz images for most members of Taurus
has been measured by \cite{luh10tau}
(see also \citealt{har05}; \citealt{luh06}; \citealt{gui07}; \citealt{reb10}).
We have measured photometry of the members that were 
not considered in \cite{luh10tau} and all known members
appearing in \spitz images that have become publicly available 
since that study (Astronomical Observation Requests 26470912, 26471168
26477056, 26475264, 26472704, 26473216, and 23272448).
These data were measured with the methods 
described by \cite{luh10tau} and are presented in Tables \ref{tab:irac} and \ref{tab:mips}.
Members identified after \cite{luh10tau}
that were not detected or observed by \spitz
are not included in these tables.

In Table \ref{tab:all},
we have constructed a compilation of all \spitz photometry from 
\cite{luh10tau} and Tables \ref{tab:irac} and \ref{tab:mips}
for all known members  of Taurus.
For members imaged at multiple epochs,
we report the mean values weighted by the inverse square of the flux errors.
Sources that lack photometry due to saturation, extended emission,
or non-detection are indicated. 
Components of binaries that are resolved by IRAC have separate entries in 
Table \ref{tab:all}.
A few of these pairs are unresolved in the MIPS images.
The combined MIPS photometry 
for these systems
is listed under the brighter component.
For the 378 members imaged by IRAC, all are detected in at least one band.
MIPS  
observed 331 members and detected 244 of them.

\subsection{\wise Photometry}\label{sec:wisephot}

In addition to the \spitz data,
we also utilize the more recent mid-IR photometry measured by {\it WISE}.
The four \wise photometric bands are centered at 3.4, 4.6, 12, and 22 $\mu$m
and are  denoted as $W1$ through $W4$ \citep{wri10}.
The first three bands have an angular resolution of $\sim$$6\arcsec$
while $W4$ has a resolution of $\sim$$12\arcsec$.
For unconfused areas near the ecliptic plane,
\wise typically achieved a signal-to-noise ratio of 5 at
$W1=16.8$, $W2=15.6$, $W3=11.3$, and $W4=8.0$ \citep{cut12}.

To compile the \wise photometry in Taurus, we began by retrieving all sources 
from the \wise All-Sky Source Catalog \citep{cut12cat} within $2\arcsec$ of
the known members.
We then inspected all images in the four \wise bands for all members that
lacked a matching \wise source.  We found that 
IRAS 04166+2706 and IRAS 04368+2557 are dominated
by extended emission in $W1$ and $W2$, which resulted in significant offsets
in the coordinates in the \wise catalog from the true positions.
HBC 360 and HBC 361 comprise an $\sim$$8\arcsec$ pair that is only partially resolved by {\it WISE}.
One source near the midpoint of the pair is present in the All-Sky Catalog but the components have separate sources associated with them in the AllWISE 
Source Catalog.
Although IRAM 04191+1522 does not have a counterpart in the All-Sky Catalog, it is visible
in the \wise images and is present in the AllWISE Source Catalog.
Therefore, we adopt the data for HBC 360, HBC 361, and IRAM 04191+1522 from the latter. 
Thirteen companions that were resolved by IRAC and hence have separate entries in our
catalog of members were unresolved from their primaries in the \wise images.

We have taken additional steps to verify the reliability of the \wise data
in the same manner preformed by \cite{luh12} for Upper Sco.
In summary, we omitted measurements of $W2$ brighter than a magnitude of 6
 because of their large systematic errors \citep{cut12},
 visually examined all \wise images for false detections,
 identified sources that may be contaminated
by extended emission or the
point-spread function (PSF) of another star,
and checked for sources whose positions differed significantly among the \wise bands. 
During the inspection of the images, we noticed that 
L1521F-IRS was detected in $W3$ and $W4$ but had a measurement in only $W4$ in the All-Sky
Catalog.
We found $W3$ photometry for it in the AllWISE Source Catalog,
which we have adopted.
For two unresolved multiple systems, HD 28867~A+B+C and
XEST 09-042+2MASS J04355949+2238291, the centroid of the WISE source shifts
with wavelength, leading us to ascribe different bands to different components
of the systems.
In total, we report \wise photometry in at least one \wise band for 
401 of the entries in Table \ref{tab:all}.

\section{Measurement of Infrared Excess Emission}
\label{sec:excess}

Circumstellar disks emit radiation predominately at IR and millimeter wavelengths.
Beyond $\sim$5 $\mu$m, this radiation surpasses photospheric emission in brightness
and can be used to detect the presence of a disk. 
Following a similar procedure to \cite{luh12},
we identify and measure excess emission of Taurus members using colors 
produced by the \spitz and \wise data 
relative to the $K$ band (2.2 $\mu m$).
In this analysis, we use the average \spitz measurements in Table \ref{tab:all}.
For most members, 
we adopt $K_s$ photometry from the 2MASS Point Source Catalog.
We use the $K$ magnitudes for 
HV Tau C, IRAS 04111+2800G, and DG Tau B %,
 from \cite{tera07},
the United Kingdom Infrared Telescope  
Infrared Deep Sky Survey \citep[UKIDSS;][]{law07},
and the 2MASS Extended Source Catalog, respectively.
We exclude the 2MASS $K_s$ magnitudes
for the binary system J1-4872 A and B because
of their large uncertainties (see \citealt{luh10tau}).
IRAS 04368+2557,  IRAM 04191+1522, 
L1521F-IRS,  IRAS 04166+2706, and SST Tau 041831.2+28161
 are known 
protostars (\citealt{fur08}; \citealt{fur11}) 
and are too heavily obscured for $K$ measurements.
We also exclude any \wise or \spitz data with errors $\ge$0.25 mag from our analysis.
As in \cite{luh12}, we examine [4.5], [8.0], [24], $W2$, $W3$, and $W4$ for excess emission.

We have corrected the \spitz and \wise colors for 
extinction prior to measuring excess emission. 
To do this, we used extinction estimates at $J$ from a variety of sources,
as indicated in Table \ref{tab:all}. These extinctions are the same as
those adopted by \citet{fur11} for the stars appearing in that study.
The values of $A_J$ were converted to color excesses with the
two reddening curves for high ($A_K > 1$) and low ($A_K \leq 1$)
extinction from \cite{mcc09}.
We did not attempt to deredden the colors for the 47 sources that lack
extinction estimates, such as protostars and edge-on disks
(see \citealt{fur11}).

\begin{figure}[h]
	\centering
	\includegraphics{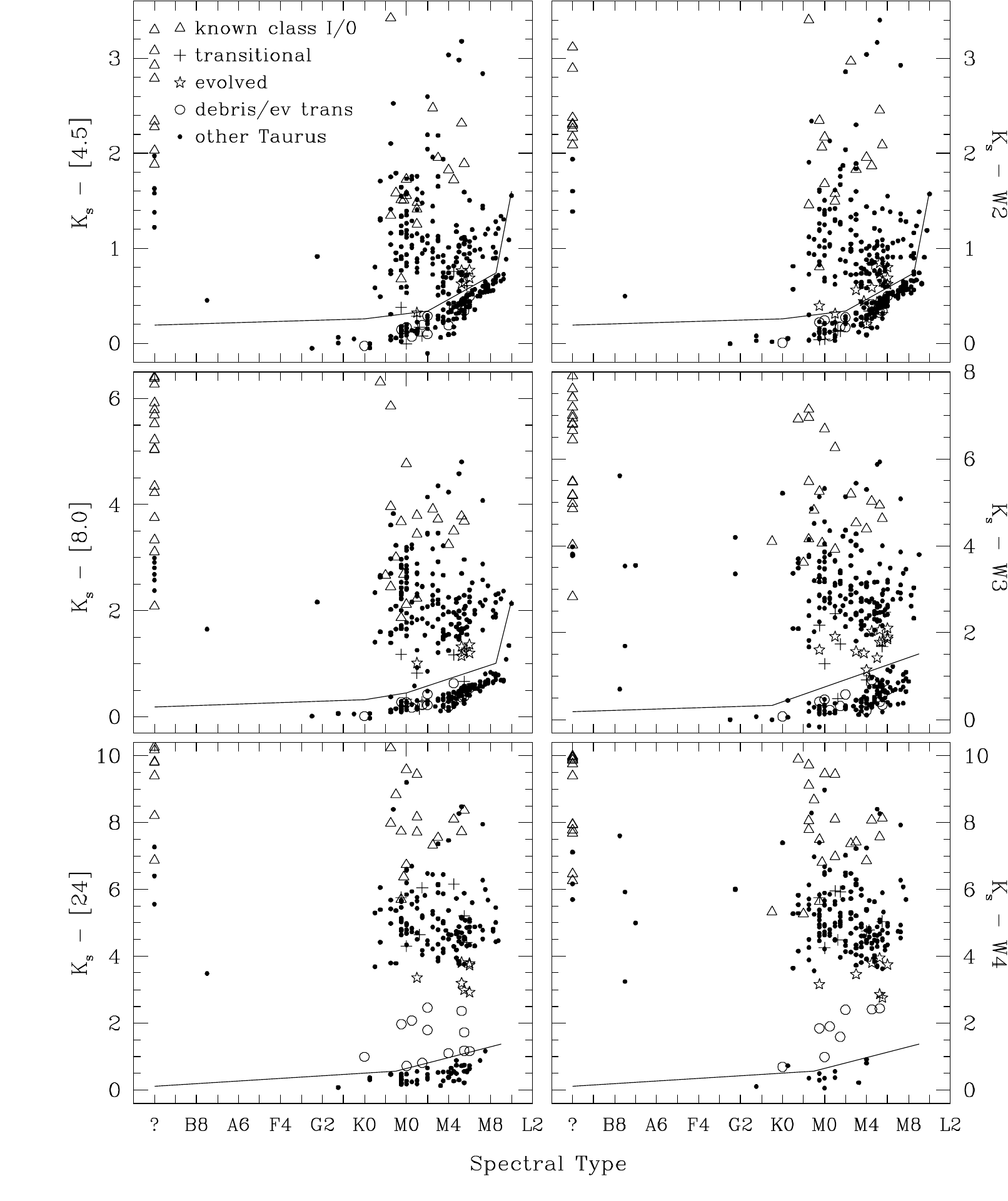}
\caption{Extinction-corrected IR colors versus spectral type for members of Taurus from \spitz (left) and \wise (right). The solid lines are used to identify the presence of excess emission from circumstellar disks. We have indicated known protostars (class I and 0, triangles), candidate transitional disks (crosses), candidate evolved disks (stars), and candidate debris disks or evolved transitional disks (circles). }
\label{fig:kvsp}
\end{figure}

Since photospheric colors vary with spectral type,
we plot the dereddened colors
 versus spectral type in Figure \ref{fig:kvsp}. 
Stars without extinction estimates are plotted at their observed colors.
For each color, 
the stellar photospheres form a narrow, blue sequence
while stars with disks have a broad distribution of redder colors.
The color thresholds used for identifying excesses are shown 
in Figure \ref{fig:kvsp}.
In the remainder of this section,
we discuss the measurements of excess emission 
in each of the six \spitz and \wise
bands that we have considered.
The bands exhibiting excess emission relative to $K$ are indicated in Table \ref{tab:all}
for each member of Taurus.
We also analyzed the excesses of members that lack $K$ photometry. 
J1-4872 A and B show no excess at longer wavelengths 
when compared to [3.6]. 
Since the remaining sources without $K$ are protostellar,  
we mark each of their detected bands as having an excess.

\subsection{Excess in [4.5] and $W2$}
 
We discuss the data for [4.5] and $W2$ together because of their similar
effective wavelengths (4.5 and 4.6 $\mu$m).
Photometry in [4.5] and $W2$ is available for 366 and 360 members of Taurus, respectively. 
Stars that lack [4.5] are either saturated, not detected (L1521F-IRS), or not imaged by IRAC. 
Members without $W2$ are too bright for good photometry, dominated by extended emission,
or blended with another object.
All of the stars imaged by IRAC without [4.5] also lack $W2$ photometry,
while 44 of the stars without $W2$ are measured in [4.5].
The 39 stars that were not observed by IRAC in [4.5]
were detected by {\it WISE}, although two of them, MWC 480 and HD 29763, are
too bright for good photometry in this band.
Consequently, 404 and 322 members have data in at least one or both of these 
two bands, respectively.

To identify the stars that exhibit significant excesses,
we have selected a boundary that follows
the sequences of stellar photospheres in the diagrams for
$K_s  - [4.5]$ and $K_s  - W2$ versus spectral type 
in Figure \ref{fig:kvsp}.
The boundary is defined by the lines connecting the points 
(B0, 0.19),
(K2, 0.26),
(M2, 0.34),
(M8.5, 0.74), and
(L0, 1.6).
The boundary is uncertain later than M8.5
because there are few members in this regime
and the colors increase rapidly at these latest spectral types.
At L0, the boundary is placed just above the colors of 
2MASS J04373705+2331080 since it lacks  
an excess at longer wavelengths \citep{luh10tau}.
Since the average offset between [4.5] and $W2$ is only $\sim$1\% \citep{cut12},
we use the same boundary for both colors. 
We also examined the objects in Figure \ref{fig:kvsp}  that lack spectral types.
All of these sources exhibit large excesses if we adopt the spectral types that are
consistent with their photometry.

We checked for discrepancies between excesses in [4.5] and $W2$ 
and data at other bands. 2MASS J04215450+2652315 and V410 Tau
have a small excess in $W2$ and $W2$/[4.5], respectively. 
However, both stars have colors consistent with stellar photospheres at longer wavelengths.
GM Aur shows excess in [4.5] and other bands
but is slightly below our boundary in $W2$.
We indicate the presence of excess in $W2$ for this star in Table \ref{tab:all}.
For all other sources with excesses in [4.5] and $W2$, 
excess emission is also detected at longer wavelengths.

\subsection{Excess in [8.0]}

Photometry at [8.0] has been measured for 369 members of Taurus.
Stars that lack [8.0] were either saturated or not observed by IRAC.
Since excess emission from disks increases at longer wavelength,
the gap between the diskless and disk-bearing members is larger 
for $K_s - [8.0]$ than it was for $K_s - [4.5]$ and $K_s - W2$,
as shown in Figure \ref{fig:kvsp}. 
We define our threshold for excess by connecting the points 
(B0, 0.19),
(K2, 0.32),
(M0, 0.45),
(M8.5, 1.01),
and (L0, 2.2).
As done in the previous section, we use 2MASS J04215450+2652315
to guide the L0 boundary. 
All objects that have excesses at [8.0] 
also exhibit excess emission at longer wavelengths when such data are available.
One star, V410 X-ray 7, has no excess at $W2$ or [4.5], a slight excess at [8.0]
and is unresolved from a nearby bright star at [24]. 
We tentatively conclude that V410 X-ray 7 has excess emission at [8.0].
The remaining stars without photometry at longer wavelengths
have fairly large excesses at [8.0] ($K_s - [8.0]$ $\gtrsim$ 1.5). 

\subsection{Excess in $W3$}
There are 358 members of Taurus with detections in $W3$
after excluding sources with errors $\ge$0.25.
Members without photometry were either unresolved from a nearby star
or not detected.
Our boundary for excess is defined by the points at
(B0, 0.18),
(G8, 0.33), and
(M9, 1.52).
As done in previous bands, 
we checked whether sources with $W3$ excesses have excesses at longer wavelengths.
2MASS J04215851+1520145 has a small excess in $W3$,
was not detected in $W4$, and was not imaged by MIPS.
The detection limit in $W4$ does not place a useful constraint on 
the presence of excess emission.
All other $W3$ excesses are large or are supported by 
detections of excess in $W4$ or [24].

\subsection{Excess in [24] and $W4$}
\label{sec:exc24}

We discuss the analysis of [24] and $W4$ together because they have 
similar effective wavelengths (23.7 and 22 $\mu$m). 
As done with $W3$, we only analyzed measurements with errors less than 0.25,
corresponding to 242 and 244 sources with [24] and $W4$, respectively.
Among the 66 sources with data in $W4$ but not [24],
45 were not imaged by MIPS,
20 were saturated in MIPS,
and one (MHO 2) was unresolved from a saturated source.
MIPS provided data for 64 members that were
not detected reliably at $W4$ ($\sigma_{\rm W4}\ge0.25$ or not detected).
Data in at least one of these two bands is available for 308 members of Taurus.

Because advanced stages of disk evolution (e.g., debris disks; see Section \ref{sec:class})
can have arbitrarily small excesses at 24 $\mu$m and no excesses at shorter wavelengths,
we have selected a boundary that attempts to identify the smallest excesses
that appear to be significant.
This boundary is defined as 
(B0, 0.11),
(K6, 0.56), and
(M9, 1.38).
Since the sequence of diskless members is not well-populated in $W4$, 
we adopt this boundary for $W4$ as well.

We find that 256 members exhibit excess emission at $W4$ or [24].
Thirteen of these stars lack excess at shorter wavelengths,
consisting of 10 stars for which [24] excesses have been noted previously 
(\citealt{fur11}; \citealt{luh10tau}, references therein) and 
new detections of excesses in $W4$ or [24] for 
2MASS J04400174+2556292,
2MASS J04414565+2301580,
and 2MASS J04242321+2650084.
The latter excesses were not identified in previous studies because
we have adopted a lower boundary for [24] excess,
the star was not imaged by MIPS,
and the object was recently added to the membership list, respectively.

Among the 106 objects without reliable photometry in either $W4$ or [24], 
86 lack excess emission at shorter wavelengths.
Seven of the 86 stars have early spectral types, and hence are bright, but they
lack good limits on $W4$ or [24] because
they are unresolved from other stars,
dominated by extended emission, 
or were not imaged by the more sensitive MIPS.
For the 79 remaining stars, which have late types, the limits on [24] and $W4$ do not place 
useful constraints
on the presence of excess emission.

Although we did not analyze any bands at wavelengths longer than [24]/$W4$,
we were able to perform two tests of our identifications of excesses.
First, we compared the results between $W4$ and [24].
Although we identified HD 286178 as a candidate member because of its $W4$ excess
(see Section \ref{sec:newmembers}),
it does not show an excess at [24]. 
HD 28929 also shows excess at $W4$ and not at [24]. 
Because the MIPS photometry is more accurate, 
we list these two stars as not having an excess in either band.
For all other stars with data at both [24] and $W4$, 
the detections of excesses agree between the two bands.
As a second test of the excesses, for the 13 stars with excess emission at
[24]/$W4$ and not at shorter wavelengths,
we examined available $70\ \mu$m photometry from the 
Photodetector Array Camera and Spectrometer (PACS; \citealt{pog10}) on the
{\it Herschel Space Observatory} \citep{pil10}.
Because the PACS images are not sensitive enough to detect the stellar photospheres of 
these stars, any detections indicate the presence of excess emission at $70\ \mu$m.
V819 Tau, JH 56, FW Tau, RXJ 0432.8+1735, 2MASS J04403979+2519061,
XEST 17-036, and 2MASS J04414565+2301580 are detected by PACS, while 
LkCa 19, V410 X-ray 3, 2MASS J04400174+2556292, and 2MASS J04242321+2650084
were imaged by PACS but were not detected \citep{how13}.
The two remaining stars, LkHa 332/G2 A+B and XEST 08-003, have not been
imaged by {\it Herschel}.
Because they are only slightly above our $W4$/[24] thresholds
 and were not detected or imaged
by {\it Herschel},
 we report only tentative detections of excesses for
 XEST 08-003, V410 X-ray 3, LkHa 332/G2 A+B, and 2MASS J04400174+2556292.

\section{Classification of Disks}
 \label{sec:class}
 \subsection{Terminology}
 
A variety of names, definitions, and classification schemes have been proposed
for the evolutionary stages of circumstellar disks.
We adopt the disk classes from \cite{esp12}, which are defined as follows:
{\it full disks} are optically thick at IR wavelengths 
and lack significant clearing of primordial dust and gas;
{\it pre-transitional} and 
{\it transitional disks} have large inner gaps or holes in their dust distributions, respectively;
{\it evolved disks} are becoming optically thin but have not experienced significant clearing;
{\it evolved transitional disks} are optically thin and have large holes;
{\it debris disks} are composed of dust generated by collisions of
planetesimals. 
Using the classification scheme described in \cite{luh12} and 
the \spitz and \wise photometry that we have complied,
we have estimated the disk classes for Taurus members that exhibit IR excess 
emission (Section \ref{sec:excess}) based on their IR colors,
as described in this section.

\subsection{Disk Classes in Taurus}
\label{subsec:disk}  
\begin{figure}[h]
	\centering
	\includegraphics{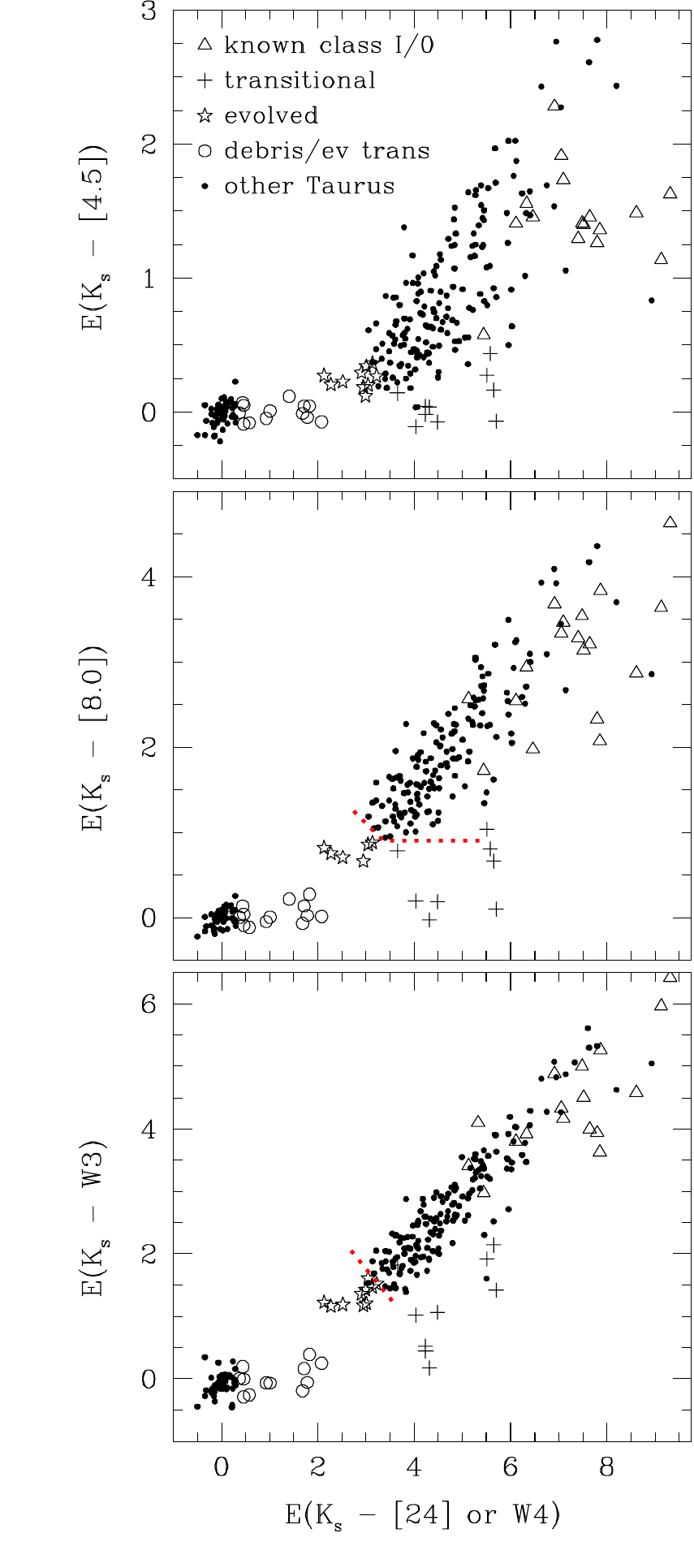}
\caption{Extinction-corrected IR color excesses for members of Taurus. Data at 4.5 and 24 $\mu m$ from \spitz are shown when available. Otherwise, measurements at similar wavelengths from \wise are used ($W2$ and $W4$). We have indicated known protostars (class I and 0, triangles), candidate transitional disks (crosses), candidate evolved disks (stars), and candidate debris disks or evolved transitional disks (circles). The reddest protostars are beyond the limits of these diagrams. 
In the middle and lower diagrams, we have marked the lower boundaries that we have adopted for full disks (dotted lines). }
\label{fig:exall}
\end{figure}

We estimated the evolutionary stages of disks using extinction-corrected color
 excesses relative to photospheric colors (e.g., $E(K_s - [24])$).
The excess in a given band was computed as the difference between the observed 
color and the average color for young stellar photospheres at the spectral
type in question. 
For the \spitz bands, we adopted the photospheric colors from \cite{luh10tau}.
The $K_s - W2$ and $K_s - W4$ colors were approximated by the similar
$K_s - [4.5]$ and $K_s - [24]$ colors, respectively.
We determined the photospheric colors for $K_s - W3$ by
 a fit to the observed diskless sequence in Figure \ref{fig:kvsp}.
The resulting excesses of Taurus members are plotted in Figure \ref{fig:exall}.
Because of their similar wavelengths, data from [4.5]/$W2$ and [24]/$W4$ 
are presented together.
When both \spitz and \wise data are available in those bands, we plot the former.
IRAS 04016+2610, Haro 6-5B, GV Tau A+B, and CoKu Tau/1 have large excesses 
in all bands and are too red to appear within the selected 
boundaries of Figure \ref{fig:exall}.

Following the procedure of \cite{luh12},
we classify the disks
 using the $K_s - [8.0]$, $K_s - W3$, and $K_s - [24]/W4$ colors.
In Figure \ref{fig:exall}, we show the boundaries 
between full disks and other sources, which are defined 
by the lines connecting the points 
($E(K_s - [24]/\mw4)$, $E(K_s - [8.0])$) = 
(2.75, 1.25), (3.36, 0.90), and (5.5, 0.90)
in the middle panel
and 
($E(K_s - [24]/\mw4)$, $E(K_s - \mw3)$) =
(2.72, 2.02) and (3.55, 1.25)
in the lower panel.
We classify disks above one or both boundaries as full.
For bluer sources, we apply the following criteria: 
transitional disks have $E(K_s - [24]/\mw4) > 3.55$;
evolved disks have $E(K_s - [24]/\mw4) < 3.55$, $E(K_s - [8.0]) > 0.3$,
and $E(K_s - \mw3) > 0.5$;
and debris disks and evolved transitional disks
have $E(K_s - [24]/\mw4) < 3.55$, $E(K_s - [8.0]) < 0.3$,
and $E(K_s - \mw3) < 0.5$.
While these boundaries reproduce the results in \cite{luh12},
they have been adjusted slightly 
to produce nearly the same classifications in Taurus as those from \cite{luh10tau}.
Because of their similar mid-IR spectral energy distributions \citep{car09}, 
debris disks and evolved transitional disks require measurements of 
gas content to be distinguished from each other.
Although GM Aur is slightly above the boundary for full disks in the middle panel of Figure \ref{fig:exall},
we mark it as transitional since it has been widely treated as such. 
Meanwhile, UX Tau A is classified as a primordial
disk because it is above the boundary in $K_s-[8.0]$, but it does have a relatively
low value of $K_s-\mw3$, which reflects the fact that it is a pre-transitional
disk \citep{fur06}. 
Although previously classified as full by \cite{luh10tau},
2MASS J04214631+2659296 is most likely a transitional disk that is becoming
optically thin because of its unusual spectral shape \citep{fur11} and
faint [24]/$W4$ photometry, which  
places it below the boundary for full disks in the second 
panel of Figure \ref{fig:exall} and above in the third panel.

We also classified the disks for members that lacked $W3$, [8.0], and/or
[24]/$W4$ but had excesses. 
Disks without [24]/$W4$ data were classified as full if the excesses at [8.0] or $W3$
were sufficiently high to exclude other disks types.
If the excesses at [8.0] and $W3$ were both too small for full disks, 
the disks were classified as evolved if the upper limits on [24]/$W4$ excluded transitional disks
and were marked as evolved or transitional if the [24]/$W4$ limits did not provide useful 
constraints.
Objects with [4.5]/$W2$ excesses but no data at longer wavelengths 
were classified as full.
Finally, all members that lack spectral classifications or $K$-band photometry (mostly protostars) 
have large enough excesses to indicate the presence of full disks.

Our disk classifications are listed in Table \ref{tab:all}.
Known or suspected protostars 
(class 0 or I) are also indicated (\citealt{fur08}; \citealt{fur11}).
The 414 entries in Table \ref{tab:all} contain 239 full disks,
10 transitional disks,
13 evolved disks,
13 evolved transitional/debris disks,
and one disk that is either evolved or transitional.
The remaining 138 members lack excess emission. 
For members known prior to this study, 
our classifications agree with those given in \cite{luh10tau} and \cite{fur11} 
except for the following: 
2MASS J04400174+2556292, 2MASS J04414565+2301580,
2MASS J04214631+2659296, 
2MASS J04390163+2336029, and 2MASS J04284263+2714039.
The first three have been discussed earlier in this study.
2MASS J04390163+2336029 lacks [4.5] and [8.0], and the data in [5.8] and [24]
considered by \cite{luh10tau} suggested a full disk. We now classify
it as an evolved disk after including the \wise bands.
2MASS J04284263+2714039 is classified as an evolved disk in this study
but not by \cite{luh10tau} because we have used extinction-corrected colors
rather than observed colors.
In addition, the new members that we have added to our census of Taurus
(see Section \ref{sec:members}) contain seven evolved disks  
(V1195 Tau, 2MASS J04380191+2519266, 2MASS J04374333+3056563,
2MASS J04215851+1520145, 2MASS J04284199+1533535, 2MASS J05073903+2311068,
and 2MASS J05122759+2253492),
three  transitional disks 
(2MASS J04355760+2253574,
2MASS J04343128+1722201, and 2MASS J05080709+2427123),
one evolved transitional/debris disk (2MASS J04242321+2650084), 
and 31 full disks.

\section{Identification of New Members of Taurus}
\label{sec:search}

\subsection{Candidate Members from \wise}

In addition to identifying and classifying the disks of known members of Taurus,
we have searched for new members that have disks via their red \wise colors.
We began by retrieving from the \wise All-Sky Source Catalog all
objects between  $4^{\rm h}00^{\rm m}$--$5^{\rm h}10^{\rm m}$
in right ascension and $15\arcdeg$--$31\arcdeg$ in declination (J2000),
which encompasses all known dark clouds and young stars in Taurus
(see Figure \ref{fig:map}).
In our analysis, we considered only data with errors less than 0.1 mag.
We excluded sources that are spurious detections of diffraction spikes in $W1$
or $W2$, as indicated by the parameter ``cc\_flag" in the WISE catalog.
Known members of Taurus were also removed from the list.
These criteria resulted in a list of $\sim$1.4 million sources.

\begin{figure}[h]
	\centering
	\includegraphics[scale=.8]{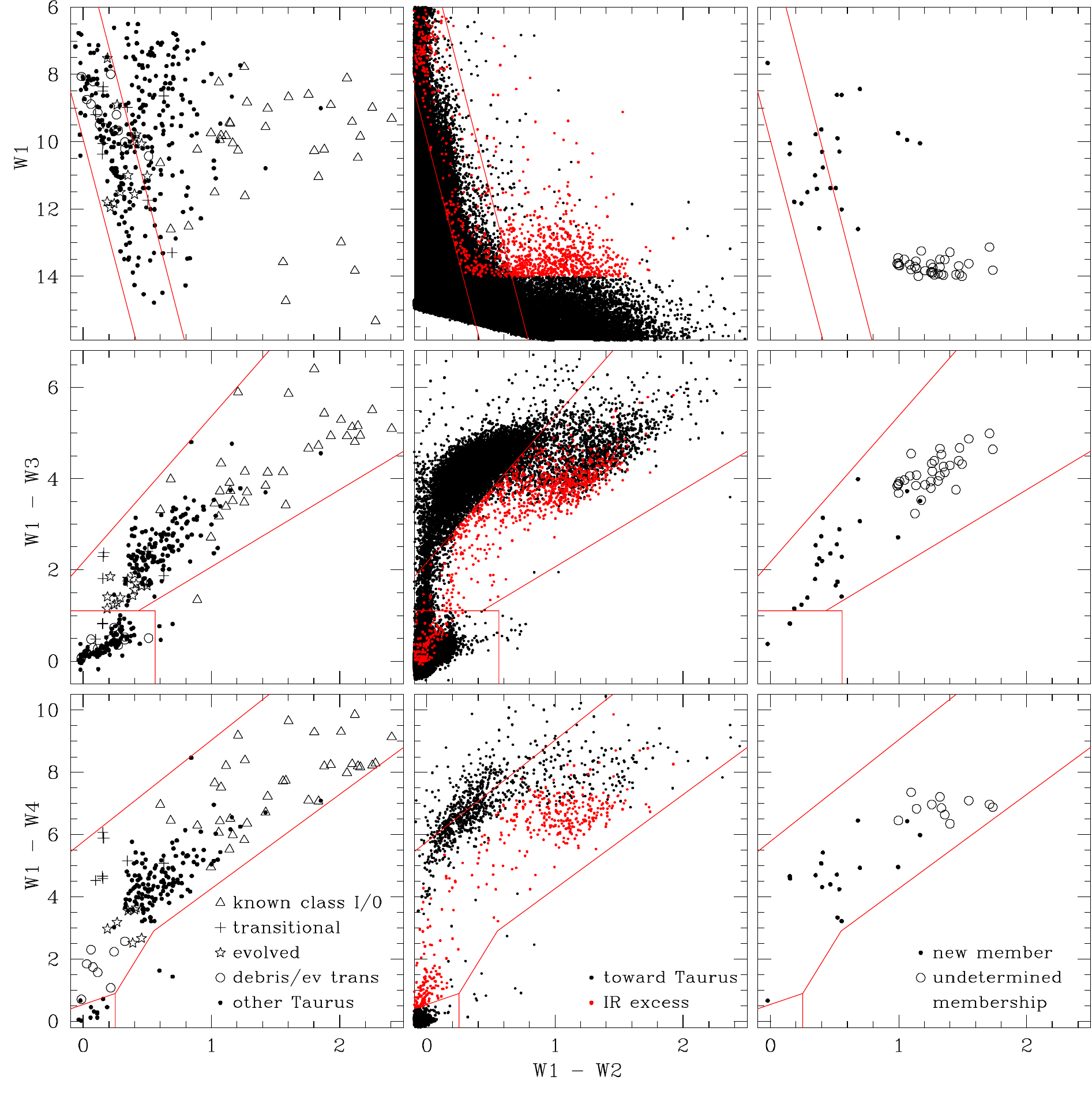}
\caption{Left: \wise color-magnitude and color-color diagrams for known members of
Taurus (same labels as Figs. \ref{fig:kvsp}-\ref{fig:exall}). We have defined regions that separately
encompass most of the members with and without excesses from disks (solid
lines).
Middle: Among \wise sources that are not known Taurus members (points),
we have identified those that have colors indicative of disks based on
the boundaries defined on the left (red points).
Right: Some of the IR excess sources from the middle diagrams
are probable non-members (Table \ref{tab:rej}). We show the positions of
the remaining candidates that have been confirmed as members through
spectroscopy (Table \ref{tab:spec}, points) and that have undetermined membership
status (Table \ref{tab:unknown}, circles). Most of the latter are probably galaxies based
on their faint magnitudes and very red colors.
}
\label{fig:candidate}
\end{figure}

We searched the catalog from \wise for objects with colors similar 
to those of known disk-bearing members of Taurus.
To demonstrate how this was done, 
we show in Figure \ref{fig:candidate} color-color and color-magnitude diagrams 
constructed from the \wise
bands for the known members of Taurus and for the other \wise sources in 
our survey field. 
To minimize contamination by non-members, we selected boundaries that cover
the smallest range of colors while also encompassing most of the known disks.
The only disk-bearing member not enclosed within these boundaries,
IRAS 04302+2247, has unusual colors because it is seen in scattered light
\citep{fur08}.
We also defined boundaries that enclose most of the members that lack excesses
and have photospheric colors.
The definitions of these boundaries are provided in Table \ref{tab:bound}.
Using these boundaries,
we identified as candidate disk-bearing members of Taurus the \wise
sources that satisfy all of the following:
1) $\mw1\le14$,
2) excess in at least one diagram,
3) excesses in all available bands long ward of some wavelength,
and 
4) in the excess or photosphere regions in all diagrams
(i.e., not in the lower right or upper left in either color-color diagram).
In addition, we rejected sources with non-detections or large errors ($\sigma >0.1$ mag)
in both $W3$ and $W4$ where 
the limits of the respective colors exclude the presence of excess emission. 
These criteria produced 1062 candidates, which
are split among Tables \ref{tab:rej}, \ref{tab:spec}, and \ref{tab:unknown},
as described in the following sections.

\subsection{Additional Membership Constraints}
\label{sec:reject}

Given that Taurus contains $\sim$250 known members with disks, it is likely
that most of the 1062 candidates from WISE are not members. To further refine
this sample of candidates, we have employed visual inspection of available
images, optical/IR color-magnitude diagrams, IRAC photometry, and 
proper motions.

We visually inspected the images of each candidate from DSS,
the Sloan Digital Sky Survey (SDSS, \citealt{yor00}), 2MASS, and {\it WISE}.
Objects were rejected if 
1) reliable detections were not present in the \wise bands that exhibited excess emission,
2) the \wise source was contaminated by emission from a nearby object,
3) the centroid shifted among the \wise bands, 
indicating that the \wise source was a blend of multiple objects,
or 
4) the candidate was resolved as a galaxy in any of the surveys. 
This inspection eliminated 454 sources.
We also rejected 39 other candidates that have been previously classified as 
galaxies, planetary nebulae, asymptotic giant branch stars, cataclysmic variables,
or other non-members according to the SIMBAD database.
The candidates \wise J043809.73+254731.5, J041810.61+284447.3 
and J041556.86+290750.9 are non-members based on unpublished spectra
from a separate survey (K. Luhman, in preparation).

\begin{figure}[h]
	\centering
	\includegraphics{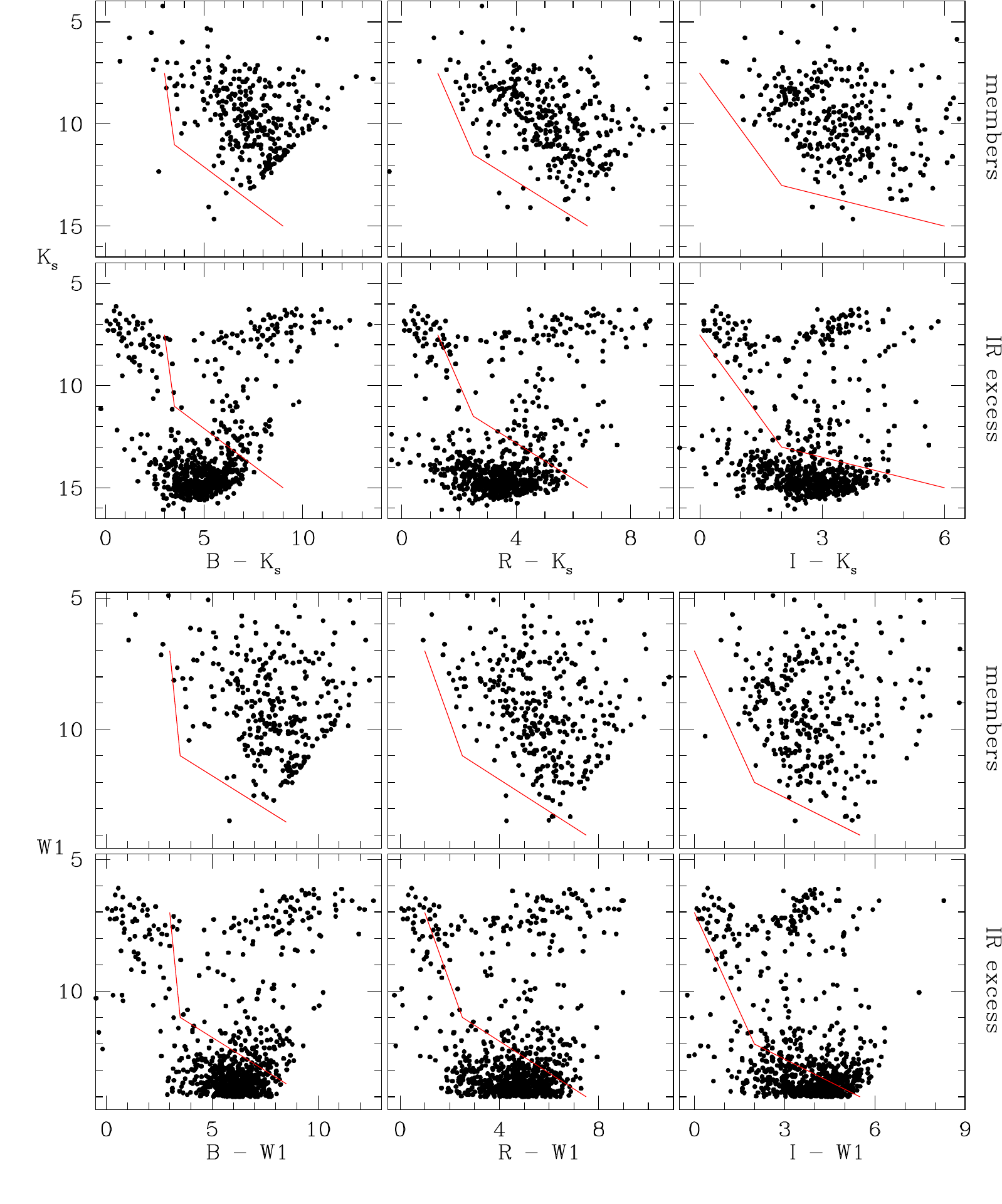}
\caption{
Color-magnitude diagrams for known members of Taurus and other {\it WISE}
sources with IR excesses
that are indicative of young stars (Figure \ref{fig:candidate}).
These diagrams are based on data from WISE ($W1$), 2MASS ($K_s$), and USNO-B1.0
(BRI). The lines indicate the lower boundaries of the population of known
members.}
\label{fig:usno}
\end{figure}

In Figure \ref{fig:usno},
we show color-magnitude diagrams constructed from 2MASS, {\it WISE}, and USNO-B1.0  
(B R I; \citealt{mon03})
for the known Taurus members and the 1062 \wise candidates. 
For objects with USNO data at two epochs, we adopted the more recent measurements. 
As done in the previous section,
we have selected boundaries in these diagrams that encompass most of the known
members, which are defined in Table \ref{tab:ubound}.
The few members that are below these boundaries are known or suspected to be
seen in scattered light (e.g., edge-on disks), 
resulting in their underluminous positions. 
For candidates detected by 2MASS with $K_s>7.5$, 
we rejected sources that are below or to the left of the boundaries in any of the
three 2MASS/USNO diagrams. 
For candidates that were not
detected by 2MASS, 
we applied the criteria from the {\it WISE}/USNO diagrams instead
($W1>7$).
In total, we reject 631 and 96 sources with the 2MASS/USNO and {\it WISE}/USNO
criteria, respectively.

Some of our \wise candidates were identified based only on red $\mw1-\mw2$ colors,
and lacked detections in $W3$ or $W4$ that could confirm the presence of excess emission.
These objects were also found to follow the spatial distribution of the Taurus dark clouds.
Since a large area in Taurus was imaged by IRAC,
we can use the available [5.8] and [8.0] data as an independent verification of excess at 
[4.5]/$W2$ for this subset of candidates.
The boundary $[5.8] - [8.0] > 0.4$ divides known members with and without
excess at [4.5]/$W2$.
Using photometry from all IRAC images of Taurus (\citealt{luh06}; \citealt{luh10tau}),
we have rejected 69 {\it WISE} candidates that lack $W3$ and $W4$ detections 
and that have $[5.8]-[8.0]<0.4$. 
Given that these candidates are projected against the dark clouds in Taurus,
they are probably background stars that are red in $\mw1-\mw2$ because of extinction.
This source of contamination should be largely eliminated from our candidate list
since IRAC imaged most of the Taurus dark clouds.

We can use proper motions to further constrain 
the membership of the \wise candidates.
Using proper motions from UCAC4 \citep{zach12},
we reject \wise candidates that differ by more than 2~$\sigma$ from
all of the average motions of the Taurus groups \citep{luh09}.
For candidates rejected by UCAC4 proper motions but
not the other criteria described above and for candidates without UCAC4 data,
we examined the proper motions from other catalogs
(\citealt{mon03}; \cite{ros08}; \citealt{roe10}; \citealt{zach10}).  
Three of these stars, 2MASS J04124068+2438157, 2MASS J05080709+2427123,
and 2MASS J05073903+2311068,
have proper motions from those catalogs that support membership,
while 2MASS J04322815+2711228 is rejected.
In addition, we examined images from DSS and 2MASS
for the 1062 candidates to check for visually discernible motions,
which would be  significantly larger than that of Taurus ($\gtrsim$100 mas/yr).
We have rejected 161 \wise candidates through these proper motion criteria.

The criteria described above rejected 976 of the 1062 
\wise candidates as likely non-members.
These objects and the criteria that they failed to satisfy are listed in
Table \ref{tab:rej}.
Although 2MASS J04051434+2008214 (HD 284154) passed all criteria, it is
rejected as a Taurus member because it and a common proper motion
companion ([WKS96] 4) are both fainter than expected for members of Taurus
with their spectral types.

\subsection{Spectroscopy of Candidate Members}
\label{sec:spectra} 

\begin{figure}[h]
	\centering
	\includegraphics[scale=.8]{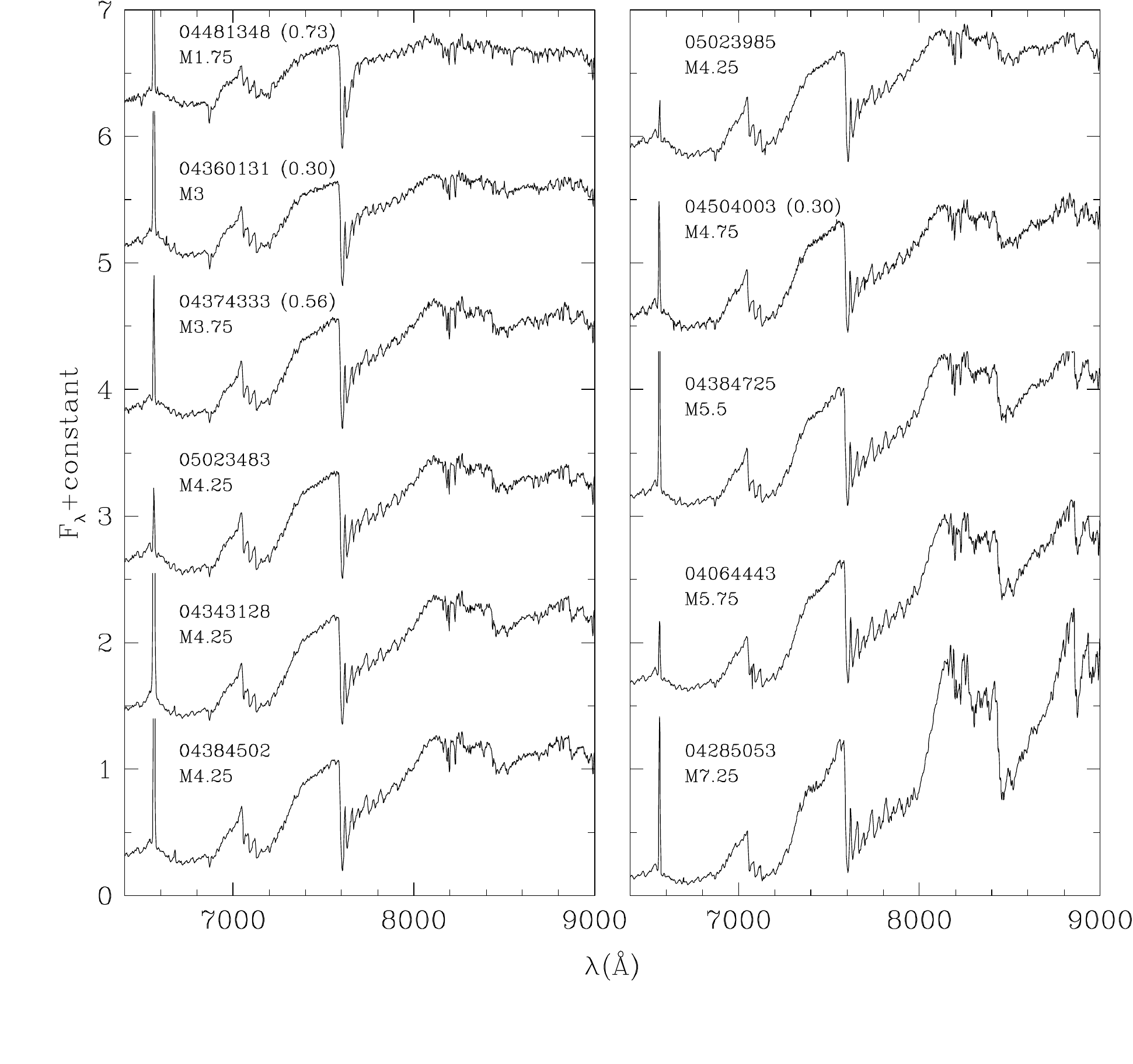}
\caption{Optical spectra of new members of Taurus. The spectra have been corrected for extinction, which is quantified in parentheses by the magnitude difference of the reddening between 0.6 and 0.9 $\mu$m ($E$(0.6-0.9)). The spectra have a resolution of 7 \AA \ and are normalized at 7500 \AA. }
\label{fig:opspec}
\end{figure}

\begin{figure}[h]
	\centering
	\includegraphics[scale=.7]{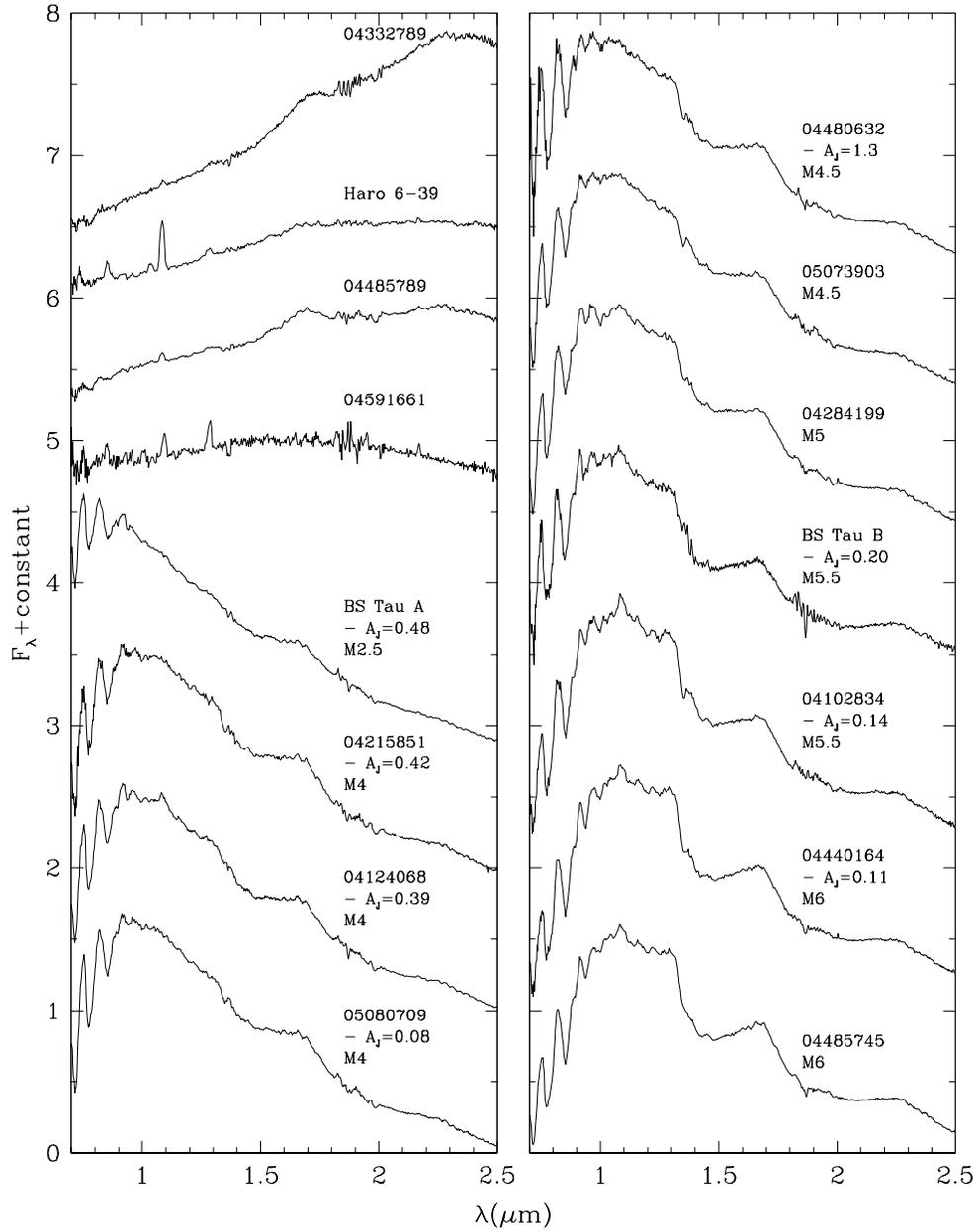}
\caption{Near-IR spectra of new members of Taurus. The spectra with measured spectral types have been corrected for extinction. These data have a resolution of $R$ = 100 and are normalized at 1.68 $\mu$m. }
\label{fig:irspec}
\end{figure}

We obtained optical and near-IR spectra of 10 and 41 \wise candidates,
respectively.
We also performed IR spectroscopy on candidate companions 
to BS Tau and 2MASS J04485789+2913548
that were noticed in their acquisition images.
Because we began spectroscopy of candidates before applying all of the membership
constraints from the previous section,
seven of the targets have positions in the USNO/2MASS/{\it WISE} color-magnitude
diagrams that are indicative of non-members.
They consist of six galaxies and a young star (2MASS J04591661+2840468).
The optical observations were performed with the
Marcario Low-Resolution Spectrograph (LRS) on the Hobby-Eberly Telescope (HET)
on the nights of 2012 December 5, 8, and 9. The instrument was operated with the G3 grism
and the $2\arcsec$ slit, which provided a wavelength coverage of
6200--9100~\AA\ and a resolution of $R=1100$.
The near-IR spectra were collected with SpeX \citep{ray03}
at the NASA Infrared Telescope Facility (IRTF) on the nights
of 2012 December 28,  2013 January 1 and  3, and 2013 August 26.
The SpeX data were collected in the prism mode 
 with a $0\farcs8$ slit,
providing a wavelength coverage of 0.8--2.5~\micron\ and a resolution
of $R=100$ for all targets except 2MASS J04221376+1525298,
which was observed with the SXD mode ($R=800$).
The data were reduced with the Spextool package \citep{cus04} and
corrected for telluric absorption \citep{vac03}.
The optical and near-IR spectra of the stars that we classify as new members
are presented in Figures \ref{fig:opspec} and \ref{fig:irspec}, respectively.

For our spectroscopic sample,
we distinguished young stars from field dwarfs by gravity-sensitive features 
(K I, Na I, H$_{2}$O) and the strength of H$\alpha$ emission.  
Galaxies were identified by their redshifted emission lines.
We measured spectral types and extinctions for the young objects 
by comparing strengths of the 
VO,  TiO and H$_2$O 
absorption bands to those of
previously known members of Taurus 
and average spectra of standard dwarfs and giants classified at optical
wavelengths \citep{luh99}. 
Our resulting classifications are presented in Table \ref{tab:spec}.
We classify 24 candidates and the companions to BS Tau and 
2MASS J04485789+2913548 as young stars, and hence likely members
of Taurus.
The remaining 27 sources that we classify as non-members are included 
in the full list of rejected candidates in Table \ref{tab:rej}.
A few of the new members have some previous evidence of youth
(e.g., BS Tau)
but have lacked spectral classifications. 
The spatial distribution of these new members is shown in Figure \ref{fig:map}.
The 34 remaining viable candidates that have not been
observed with spectroscopy are presented in Table \ref{tab:unknown}.

We have additional comments on the classifications of some of the objects
in our spectroscopic sample.
BS Tau is a 1.3\arcsec\ binary with spectral types M2.5 and M5.5. The primary
appears to be the source of the \wise excess based on a comparison of 
astrometry in 2MASS and \wise and its strong H$\alpha$ emission,
which is a signature of accretion.
2MASS J04332789+1758436, 2MASS J04485789+2913548, and 
Haro 6-39 are very red and show H emission lines. They are not fit well by
any standards and probably have both blue and red excesses.
2MASS J04591661+2840468 may be seen in scattered light since it is very
faint for its color (e.g., edge-on disk).
Our near-IR spectrum of 2MASS J04221376+1525298 is indicative of a 
early F star with a reddening that corresponds to A$_V\sim4$ and significant
$K$-band continuum veiling. The latter combined with the \wise excess emission
suggest that it has a disk, and hence is a young star.
However, it is fainter than expected for an F-type member of Taurus,
indicating that it is seen in scattered light or is a background young star.
We classify it as the latter for the purposes of this study.
We note that this star also shows strong absorption in He I at 1.083 $\mu$m
($W_{\lambda}\sim5$~\AA), which is unusual for an F star and may be due
to a wind \citep{edw03}.
Among the seven M dwarfs that we classify as non-members, five have only
marginal excesses in a single band and thus probably do not have real excesses.
The remaining two M dwarfs, 
2MASS J04152336+3006258 and 2MASS J04503102+1514127,
seem to have significant excesses and are much fainter than typical members at their 
spectral types. 
Since these two stars do not show any evidence of being edge-on disks, 
they are probably field M dwarfs that have debris disks or that are unresolved 
from background red galaxies.

\subsection{Other New Members}
\label{sec:newmembers}

While searching for new disk-bearing members based on red \wise colors,
we found that four excess candidates 
already had sufficient evidence of membership
from previous work for inclusion in our initial sample of members
(Section \ref{sec:members}).
These stars consist of V1195 Tau, HD 31305, RXJ 0432.7+1809, and HD 286178.
V1195 Tau is a 1\arcsec\ binary \citep{koh98} that is somewhat isolated
from known members.
Its proper motion \citep{ros08} and radial velocity \citep{wic00} are consistent with membership
\citep{luh09}.
Since \cite{wic00} did not detect Li,
V1195 Tau has generally been considered a background star.
However, it was detected by \cite{mar99} and \cite{ngu12}.
HD 31305 is close to known members, is an X-ray source \citep{wic96},
and has a proper motion \citep{zach12}
that agrees with the closest subgroup \citep{luh09}.
Its photometry is consistent with a A0 zero age main sequence (ZAMS) star at the
distance of Taurus, which makes it the second most massive/hottest star in that 
group behind  AB Aur. 
We estimate that its position on the Hertzsprung-Russell (H-R) diagram
(log(T$_{eff}$) $\simeq$ 3.96,
log(L/L$_{\odot}$) $\simeq$ 1.32) is consistent with being somewhat
older (5-10 Myr) than AB Aur \citep[1.5 Myr;][]{Palla02}, but similar
in age to MWC 480 \citep[7.4 Myr;][]{Manoj06}. 
\cite{cod13} and \cite{moo13}\footnote{We reject the other four stars identified by 
\cite{moo13} as members of Taurus (See Appendix).}
identified HD 31305 as a probable member
of Taurus based on variability and proper motion, respectively.
RXJ 0432.7+1809 and HD 286178
 were classified by \cite{wic96} as young stars 
because of their X-ray emission and Li absorption. 
In addition, \cite{ses08} confirmed the Li detection for HD 286178.
The proper motions of these stars are consistent with membership in Taurus
(\citealt{ros08}; \citealt{zach12}).
We identified HD 286178 as a candidate because the criteria in Figure 3 suggested it
had an excess at $W4$. 
However, the more detailed analysis that we applied to the known members
indicates that it probably does not have an excess.

\subsection{Comparison to Rebull et al.\ 2011}

We compare the results of our survey for new disk-bearing members of Taurus
with the {\it WISE}-based search performed by \cite{reb11}.
That study recovered 18 candidate members originally found by \cite{reb10}
and identified 94 additional candidates.
Among those 112 candidates,
39 are rejected through the criteria in Section \ref{sec:reject},
12 are beyond the limit of our survey ($\mw1 > 14$),
23 do not satisfy our criteria for excess emission,
26 are outside our search area,
one is the member V1195 Tau (see Section \ref{sec:newmembers}),
and 11 are included in our spectroscopic sample.
In this last group, one is rejected (2MASS J04221376+1525298)
and ten are confirmed as members by our spectra.
Meanwhile, four of our new members are among the stars rejected
by \cite{reb11}, 
which are 
2MASS J04343128+1722201,
2MASS J04485789+2913548, 
2MASS J04591661+2840468, 
and 2MASS J05080709+2427123.
They also described HD~31305 as a likely field star,
which we classify as a member (Section \ref{sec:newmembers}).
Twenty-five of our candidates in Table \ref{tab:unknown} were
examined by \cite{reb11}, who classified them as likely galaxies based on their
spectral energy distributions.
The remaining 9 candidates 
were not considered in \cite{reb11}. 
They also did not evaluate eight of our spectroscopically confirmed members.
Finally, the list of known members from \cite{reb11} 
includes two stars, St H$\alpha$ 34 and IRAS 04262+2735,
that we have omitted as non-members 
(\citealt{har05b}; \citealt{luh09})\footnote{\cite{luh09}
spectroscopically classified IRAS 0462+2735 as an M
giant. Its proper motion is also inconsistent with membership in Taurus.}.

\subsection{Completeness of Census of Members with Disks}

We now evaluate the completeness of the current census of Taurus for
disk-bearing members.
The {\it WISE} catalog is $\sim95$\% complete at $W1=16.9$ and $W2=15.5$
for uncrowded areas with the depth of coverage found in Taurus \citep{cut12}.
Meanwhile, our criteria in Table \ref{tab:bound} encompass nearly all known
members of Taurus with mid-IR excesses.
Therefore, our sample of candidates for new disks in Taurus should have a
high level of completeness down to our adopted limit of $W1=14$.
The magnitudes of the 34 candidates with uncertain membership status are
indicated in Figure \ref{fig:candidate}.
There are no candidates with $\mw1 < 13$, indicating  
that the census of known Taurus members with IR excess should be largely complete
in this magnitude range, with the exception of objects that
are unresolved from brighter stars.
Among the 26 known members of Taurus from M7 to M8, 
the faintest $W1$ measurement is 13.05.
Thus, our census of disk-bearing members should be complete to $\sim$M8. 
The completeness is lower at fainter levels, but most of these candidates are
probably galaxies based on their very red colors \citep{reb11}, although a
few may be protostars.

\section{Conclusions}

We have performed a survey of circumstellar disks in the Taurus star-forming
region using {\it WISE}. 
Our results are summarized as follows:

\begin{enumerate}

\item 
We have examined all images from the cryogenic phase of \wise for all known
members of Taurus to check  for false detections, 
blends with nearby objects, 
and extended emission.
We have presented the resulting catalog of vetted photometry in the \wise bands
at 3.4, 4.6, 12, and 22 $\mu$m ($W1$--$W4$).
All resolved, unblended members are detected by \wise in at least one band.

\item By using colors constructed from $K_s$ and six \spitz and \wise bands,
we have identified Taurus members showing excess emission from circumstellar disks
and have estimated the evolutionary stages of the detected disks, consisting 
of full, transitional, evolved, evolved transitional, and debris disks.
Our classifications are generally consistent with those found by \cite{luh10tau}
and \cite{fur11}.
We have found 31 new full disks and 16 new candidate disks in the more 
advanced evolutionary stages.

\item Using photometry from {\it WISE}, {\it Spitzer}, 2MASS, and USNO, proper motions from UCAC4 and other catalogs, and our optical and near-IR spectroscopy,
we have found 26 new members of Taurus with spectral types of M1 -- M7.
The census of disk-bearing stars in Taurus should now be largely complete for $\mw1<13$
($\lesssim$M8;  $M \gtrsim 0.03$~$M_\odot$).

\end{enumerate}

\acknowledgements
T.E. and K.L. were supported by grant NNX12AI58G from the NASA Astrophysics
Data Analysis Program.
E.E.M. acknowledges support from NSF grants AST-1008908 and AST-1313029.
This publication makes use of data products from the Wide-field Infrared Survey Explorer, which is a joint project of the University of California, Los Angeles, and the Jet Propulsion Laboratory/California Institute of Technology, and NEOWISE, which is a project of the Jet Propulsion Laboratory/California Institute of Technology. WISE and NEOWISE are funded by the National Aeronautics and Space Administration (NASA).
The {\it Spitzer Space Telescope} and the IPAC Infrared Science Archive (IRSA)
are operated by JPL and Caltech under contract with NASA.
2MASS is a joint project of the University of
Massachusetts and the Infrared Processing and Analysis Center (IPAC) at
Caltech, funded by NASA and the NSF.
The IRTF is operated by the University of Hawaii under cooperative agreement
NNX-08AE38A with NASA.
The HET is a joint project of the University of Texas at Austin, the
Pennsylvania State University, Stanford University,
Ludwig-Maximillians-Universit\"at M\"unchen, and Georg-August-Universit\"at
G\"ottingen and is named in honor of its principal benefactors, William
P. Hobby and Robert E. Eberly.
The Marcario Low-Resolution Spectrograph at HET is named for Mike Marcario
of High Lonesome Optics, who fabricated several optics for the
instrument but died before its completion; it is a joint project of
the HET partnership and the Instituto de Astronom\'{\i}a de la
Universidad Nacional Aut\'onoma de M\'exico.
The Digitized Sky Survey was produced at the Space Telescope Science
Institute under U.S. Government grant NAG W-2166. The images of these
surveys are based on photographic data obtained using the Oschin Schmidt
Telescope on Palomar Mountain and the UK Schmidt Telescope. The plates
were processed into the present compressed digital form with the
permission of these institutions.
The Center for Exoplanets and Habitable Worlds is supported by the
Pennsylvania State University, the Eberly College of Science, and the
Pennsylvania Space Grant Consortium.

\appendix
\section{Comments on B/A stars from Mooley et.\ al (2013)}
From among the five new proposed B/A-type Taurus candidates from
\cite{moo13}, we retain only HD 31305 as a probable Taurus
member (see \ref{sec:newmembers}) and reject the other four stars
(HD 28929, $\tau$ Tau, 72 Tau, and HD 26212). Below, we describe our reasoning
for excluding the latter stars from our membership sample.

{\it HD 28929} is a B8V star \citep{ken94}
with a common proper motion companion, HD~28929B
(2MASS J04343987+2857347, UCAC4 595-013167, TYC 1841-1391-1).
The BVJHK$_s$ photometry for HD 28929B is similar to that of a F6V
star \citep{ofek08}, and so if it has the same reddening as HD~28929A
(E(B-V) $\simeq$ 0.05), then it is likely a mid-F star. 
If the revised Hipparcos parallax for HD 28929A ($\varpi$ =
6.72\,$\pm$\,0.34 mas) is adopted, then the companion has M$_V$ $\simeq$ 4.2,
which is more consistent with a main sequence star rather than a
pre-main sequence star. \citet{Zorec12} estimates an isochronal age of
HD 28929 of $\sim$600 Myr. Using the estimated H-R diagram position for
HD 28929 A of T$_{eff}$ = 12850 K and log(L/L$_{\odot}$) $\simeq$
2.30, we estimate an age of $\sim$120 Myr using the \citet{Bertelli09}
evolutionary tracks. Although the tangential motion of HD 28929 differs from
that of the nearest Taurus subgroups at a level of only $\sim$2-3 km/s, and
it appears to be co-distant with Taurus, the HD 28929 system is isolated and
has no other known Taurus members within a degree.
Based on all of these considerations, and the lack of
indicators of extreme youth for either component, we conclude that the HD 28929
system is likely to be a $\sim$120 Myr-old interloper.

{\it $\tau$ Tau} (HD 29763) is a B3V triple system \citep{les68}. 
We disagree with the assessment of \citet{moo13} that the star appears to be a
 kinematic match to the Tau V subgroup, which is in its vicinity. Combining the
 systemic velocity from \citet{Petrie61} (12.3 km/s) with the revised
 Hipparcos astrometry, we estimate its space motion to be (U, V, W)
 $\simeq$ -10, -8, -12 km/s, which is not near that of the Taurus
 subgroups \citep[see Table 8 of][]{luh09}. The star's tangential
 motion differs by $\sim$4-6 km/s from the mean motion of the nearest Tau
 subgroups (IV and V), with its proper motion in right ascension going the wrong
 direction, and the systemic radial velocity differs by $\sim$4 km/s from
 that of Taurus. $\tau$ Tau A is not near the ZAMS
 (log(T$_{eff}$) $\simeq$ 4.20\,$\pm$\,0.01 dex, log(L/L$_{bol}$) $\simeq$
 2.9\,$\pm$\,0.1 dex); indeed, we estimate an isochronal age of
 $\sim$60 Myr using the \citet{Bertelli09} tracks. $\tau$ Tau B is a
 A1V star \citep{Tolbert64}, for which we estimate $M_V$ $\simeq$ 1.6.
 As a $\sim$2.1 M$_{\odot}$ main sequence star, it would take a star
 of $\tau$ Tau B's mass $\sim$8 Myr to contract to reach the main
 sequence, which sets a strong lower limit on the age of the $\tau$
 Tau system, and rules out the possibility that $\tau$ Tau A might be
 in the pre-main sequence phase.
We conclude that the $\tau$ Tau system is a $\sim$60 Myr-old interloper.

{\it 72 Tau} (HD 28149) is a B7V star \citep{les68} near the Tau~V subgroup. Its
tangential motion differs by $\sim$6 km/s from that of
the Tau~V group \citep{luh09}. The radial velocity for 72 Tau
(7.3\,$\pm$\,2.6 km/s) is significantly different from that of Tau~V
(15.7 km/s). Its space velocity \citep[U, V, W = -5.9, -5.3, -7.4
km/s;][]{Anderson12} differs from that of the Tau V subgroup
by 11~km/s \citep{luh09}.  \citet{Zorec12} estimates an isochronal age of
179\,$\pm$\,87 Myr, although our estimated H-R diagram position for 72
Tau (log(T$_{eff}$) $\simeq$ 4.16, log(L/L$_{\odot}$) $\simeq$ 2.32)
places the star closer to the ZAMS ($<$15 Myr?). Despite its youth, we
find that 72 Tau is a poor kinematic match for Taurus membership, and
consider it an interloper.

{\it HD 26212} is a A5V star \citep{Grenier99, moo13} at a distance of
100\,$\pm$\,7 pc \citep{vanLeeuwen07}, placing it $>$5$\sigma$ closer than
the mean distance to Taurus (140 pc). HD 26212 is not in the vicinity
of any of the Taurus subgroups ($>$2 deg from Tau VIII), and its
velocity \citep[U, V, W = -18, -5, -12 km/s;][]{Anderson12} is not a
good match to the Taurus subgroups \citep{luh09}.  We estimate
that the star is lightly reddened (E(B-V) = 0.04\,$\pm$\,0.02) with an H-R
diagram position of log(T$_{eff}$) = 3.907\,$\pm$\,0.007 dex,
log(L/L$_{\odot}$) = 1.01\,$\pm$\,0.07 dex. Comparing this position to
pre-main-sequence isochrones and members of Upper Sco in Figure~14 of
\citet{Pecaut12}, it appears that HD 26212 is a ZAMS star and is below the
Upper Sco ($\sim$11 Myr) A-star sequence. Hence, it is almost
certainly $>$10 Myr, and not a pre-main-sequence star.  Given the
discordance of its position, distance, velocity, and age compared to
Taurus, and lack of secondary youth indicators (e.g., IR excess), we
consider HD 26212 an interloper.

\clearpage

\begin{deluxetable}{ll}
\tabletypesize{\scriptsize}
\tablewidth{0pt}
\tablecaption{\spitz and \wise Data for Members of Taurus\label{tab:all}}
\tablehead{
\colhead{Column Label} &
\colhead{Description}}
\startdata
2MASS & Source name from the 2MASS Point Source Catalog \\
WISE & Source name from the WISE All-Sky Source Catalog\tablenotemark{a} \\
OtherNames & Other source names \\
SpType & Spectral type \\
Aj & J-band extinction \\
f\_Aj & Method of extinction estimation\tablenotemark{b} \\
3.6mag & {\it Spitzer} [3.6] band magnitude \\
e\_3.6mag & Error in [3.6] band magnitude \\
f\_3.6mag & Flag on [3.6] band magnitude\tablenotemark{c} \\
4.5mag & {\it Spitzer} [4.5] band magnitude \\
e\_4.5mag & Error in [4.5] band magnitude \\
f\_4.5mag & Flag on [4.5] band magnitude\tablenotemark{c} \\
5.8mag & {\it Spitzer} [5.8] band magnitude \\
e\_5.8mag & Error in [5.8] band magnitude \\
f\_5.8mag & Flag on [5.8] band magnitude\tablenotemark{c} \\
8.0mag & {\it Spitzer} [8.0] band magnitude \\
e\_8.0mag & Error in [8.0] band magnitude \\
f\_8.0mag & Flag on [8.0] band magnitude\tablenotemark{c} \\
24mag & {\it Spitzer} [24] band magnitude \\
e\_24mag & Error in [24] band magnitude \\
f\_24mag & Flag on [24] band magnitude\tablenotemark{c} \\
W1mag & {\it WISE} $W1$ band magnitude\\
e\_W1mag & Error in $W1$ band magnitude \\
f\_W1mag & Flag on $W1$ band magnitude\tablenotemark{c} \\
W2mag & {\it WISE} $W2$ band magnitude \\
e\_W2mag & Error in $W2$ band magnitude \\
f\_W2mag & Flag on $W2$ band magnitude\tablenotemark{c} \\
W3mag & {\it WISE} $W3$ band magnitude\\
e\_W3mag & Error in $W3$ band magnitude \\
f\_W3mag & Flag on $W3$ band magnitude\tablenotemark{c} \\
W4mag & {\it WISE} $W4$ band magnitude\\
e\_W4mag & Error in $W4$ band magnitude \\
f\_W4mag & Flag on $W4$ band magnitude\tablenotemark{c} \\
Exc4.5 & Excess present in [4.5]?  \\
Exc8.0 & Excess present in [8.0]? \\
Exc24 & Excess present in [24]? \\
ExcW2 & Excess present in $W2$? \\
ExcW3 & Excess present in $W3$? \\
ExcW4 & Excess present in $W4$? \\
DiskType & Disk type \\
\enddata
\tablecomments{The table is available in a machine-readable format.}
\tablenotetext{a}{Source names for HBC 360, HBC 361, and IRAM 04191+1522
are from the ALLWISE Source Catalog.}
\tablenotetext{b}{
$J-H$ and $J-K$= derived from these 2MASS colors assuming photospheric
near-infrared colors;
CTTS = derived from $J-H$ and $H-K$ colors assuming intrinsic CTTS colors 
from \cite{mey97};
opt. spec. = derived from an optical spectrum;
SpeX = derived from SpeX spectrum;
(1) = \cite{bri98};
(2) = \cite{luh00};
(3) = \cite{ss94};
(4) = \cite{bec07};
(5) = \cite{whi01};
(6) = \cite{dew03};
(7) = \cite{cal04}.
}
\tablenotetext{c}{
nodet = non-detection;
sat = saturated;
out = outside of the camera's field of view;
bl = photometry may be affected by blending with a nearby star;
ext = photometry is known or suspected to be contaminated by extended emission
(no data given when extended emission dominates);
bin = includes an unresolved binary companion;
unres = too close to a brighter star to be detected;
false = detection from {\it WISE} catalog that appears false or unreliable
based on visual inspection;
err = $W2$ magnitudes brighter than $\sim$6~mag are erroneous.}
\end{deluxetable}

\clearpage

\begin{deluxetable}{llrrrrl}
\tabletypesize{\scriptsize}
\tablewidth{0pt}
\tablecaption{New IRAC Photometry for Members of Taurus\label{tab:irac}}
\tablehead{
\colhead{2MASS\tablenotemark{a}} &
\colhead{Name} &
\colhead{[3.6]} &
\colhead{[4.5]} &
\colhead{[5.8]} &
\colhead{[8.0]} &
\colhead{Date}
}
\startdata
J04034930+2610520 &         HBC 358 A+B+C   &  9.15$\pm$0.02 &        out &  8.99$\pm$0.03 &     out & 2009 Mar 20 \\
J04034997+2620382 &      XEST 06-006   & 11.99$\pm$0.02 & 11.89$\pm$0.02 & 11.81$\pm$0.03 & 11.76$\pm$0.04 & 2009 Mar 20 \\
J04035084+2610531 &          HBC 359   &  9.29$\pm$0.02 &        out &  9.20$\pm$0.03 &     out & 2009 Mar 20 \\
J04105425+2501266 &    \nodata & 11.90$\pm$0.02 & 11.32$\pm$0.02 & 10.92$\pm$0.03 & 10.38$\pm$0.03 & 2005 Feb 20 \\
J04144158+2809583 &    \nodata & 14.59$\pm$0.03 & 14.26$\pm$0.03 & 14.04$\pm$0.08 & 14.30$\pm$0.16 & 2005 Feb 19 \\
                  &            & 14.56$\pm$0.03 & 14.39$\pm$0.03 & 13.96$\pm$0.09 & 13.76$\pm$0.13 & 2007 Mar 29 \\
J04153452+2913469 &    \nodata & 12.27$\pm$0.02 & 10.95$\pm$0.02 &  9.99$\pm$0.03 &  9.01$\pm$0.03 & 2007 Mar 29 \\
J04153566+2847417 &    \nodata & 13.14$\pm$0.02 & 12.21$\pm$0.02 & 11.31$\pm$0.03 & 10.36$\pm$0.03 & 2007 Mar 29 \\
J04154131+2915078 &    \nodata & 11.24$\pm$0.02 & 11.10$\pm$0.02 & 11.05$\pm$0.03 & 10.96$\pm$0.03 & 2007 Mar 29 \\
J04154269+2909558 &    \nodata & 10.84$\pm$0.10 & 10.76$\pm$0.10 & 10.67$\pm$0.10 & 10.54$\pm$0.10 & 2007 Mar 29 \\
J04154807+2911331 &    \nodata & 13.47$\pm$0.02 & 13.34$\pm$0.03 & 13.15$\pm$0.05 & 13.28$\pm$0.10 & 2007 Mar 29 \\
J04161726+2817128 &    \nodata &  9.42$\pm$0.02 &  9.30$\pm$0.02 &  9.22$\pm$0.03 &  9.28$\pm$0.03 & 2005 Feb 19 \\
                  &            &  9.41$\pm$0.02 &  9.33$\pm$0.02 &  9.27$\pm$0.03 &  9.26$\pm$0.03 & 2007 Mar 29 \\
          \nodata &  SST Tau 041831.2+282617  & 14.10$\pm$0.03 & 13.35$\pm$0.02 & 12.82$\pm$0.04 & 11.98$\pm$0.04 & 2005 Feb 19 \\
                  &            & 14.01$\pm$0.02 & 13.18$\pm$0.02 & 12.62$\pm$0.04 & 11.80$\pm$0.04 & 2005 Feb 21 \\
                  &            & 14.07$\pm$0.03 &        out & 12.62$\pm$0.05 &     out & 2007 Mar 29 \\
J04213965+2649143 &    \nodata & 11.09$\pm$0.02 & 10.95$\pm$0.02 & 10.96$\pm$0.03 & 10.85$\pm$0.03 & 2007 Oct 17 \\
J04215482+2642372 &    \nodata & 10.85$\pm$0.02 & 10.69$\pm$0.02 & 10.66$\pm$0.03 & 10.68$\pm$0.03 & 2007 Oct 17 \\
J04242321+2650084 &    \nodata &  9.42$\pm$0.02 &  9.35$\pm$0.02 &  9.27$\pm$0.03 &  9.20$\pm$0.03 & 2005 Feb 21 \\
J04245021+2641006 &    \nodata & 10.91$\pm$0.02 & 10.77$\pm$0.02 & 10.73$\pm$0.03 & 10.75$\pm$0.03 & 2004 Sep 7 \\
                  &            & 10.90$\pm$0.02 & 10.78$\pm$0.02 & 10.66$\pm$0.03 & 10.73$\pm$0.03 & 2005 Feb 21 \\
J04251550+2829275 &    \nodata &  9.91$\pm$0.02 &  9.76$\pm$0.02 &  9.69$\pm$0.03 &  9.72$\pm$0.03 & 2005 Feb 22 \\
J04264449+2756433 &    \nodata & 11.19$\pm$0.02 & 11.19$\pm$0.02 & 11.01$\pm$0.03 & 11.01$\pm$0.03 & 2005 Feb 22 \\
J04272467+2624199 &    \nodata &  9.92$\pm$0.02 &  9.85$\pm$0.02 &  9.85$\pm$0.03 &  9.82$\pm$0.03 & 2005 Feb 22 \\
J04285053+1844361 &    \nodata &        out &  9.42$\pm$0.02 &        out &  8.38$\pm$0.03 & 2004 Mar 8 \\
                  &            &  9.70$\pm$0.02 &  9.41$\pm$0.02 &        out &     out & 2011 Nov 29 \\
J04314644+2506236 &    \nodata & 10.70$\pm$0.02 & 10.72$\pm$0.02 & 10.55$\pm$0.03 & 10.58$\pm$0.03 & 2005 Feb 24 \\
J04324107+1809239 &  RXJ 0432.7+1809  &        out &  9.55$\pm$0.02 &        out &  8.82$\pm$0.03 & 2005 Feb 19 \\
J04325323+1735337 &  RXJ 0432.8+1735  &  8.91$\pm$0.02 &  8.89$\pm$0.02 &  8.79$\pm$0.03 &  8.75$\pm$0.03 & 2004 Sep 7 \\
J04332789+1758436 &    \nodata &        out &  9.31$\pm$0.02 &        out &  7.77$\pm$0.03 & 2005 Feb 19 \\
                  &            &        out &  9.06$\pm$0.02 &        out &     out & 2011 Nov 10 \\
J04333278+1800436 &    \nodata &        out &  8.90$\pm$0.02 &        out &     out & 2011 Nov 10 \\
J04334871+1810099 &           DM Tau    &  9.38$\pm$0.02 &  9.17$\pm$0.02 &        out &     out & 2011 Nov 10 \\
J04340619+2418508 &    \nodata &        out & 12.36$\pm$0.02 &        out & 12.29$\pm$0.04 & 2004 Mar 12 \\
                  &            & 12.48$\pm$0.02 & 12.39$\pm$0.02 & 12.51$\pm$0.04 & 12.34$\pm$0.04 & 2005 Feb 24 \\
J04345973+2807017 &    \nodata & 14.34$\pm$0.03 & 13.95$\pm$0.03 & 13.71$\pm$0.06 & 13.15$\pm$0.07 & 2005 Feb 20 \\
                  &            & 14.21$\pm$0.03 & 13.91$\pm$0.03 & 13.79$\pm$0.07 & 13.04$\pm$0.07 & 2007 Oct 16 \\
J04355760+2253574 &    \nodata & 12.86$\pm$0.02 & 12.46$\pm$0.02 & 12.19$\pm$0.03 & 12.15$\pm$0.04 & 2005 Feb 20 \\
                  &            & 12.86$\pm$0.02 &        out & 12.34$\pm$0.04 &     out & 2005 Feb 21 \\
                  &            & 13.27$\pm$0.02 & 12.84$\pm$0.02 & 12.52$\pm$0.04 & 12.30$\pm$0.05 & 2007 Apr 3 \\
J04355881+2438404 &    \nodata &  9.57$\pm$0.02 &        out &  9.50$\pm$0.03 &     out & 2004 Mar 11 \\
                  &            &  9.55$\pm$0.02 &  9.48$\pm$0.02 &  9.45$\pm$0.03 &  9.49$\pm$0.03 & 2005 Feb 21 \\
J04355949+2238291 &    \nodata & 12.34$\pm$0.03 & 11.64$\pm$0.03 & 10.83$\pm$0.03 &  9.60$\pm$0.03 & 2007 Apr 3 \\
J04363248+2421395 &    \nodata & 10.08$\pm$0.02 &  9.99$\pm$0.02 &  9.97$\pm$0.03 &  9.90$\pm$0.03 & 2005 Feb 21 \\
J04380191+2519266 &    \nodata & 11.79$\pm$0.02 & 11.72$\pm$0.02 & 11.56$\pm$0.03 & 10.99$\pm$0.03 & 2005 Feb 21 \\
J04383907+1546137 &       HD 285957   &  8.19$\pm$0.02 &  8.23$\pm$0.02 &  8.20$\pm$0.03 &  8.20$\pm$0.03 & 2004 Sep 7 \\
J04504003+1619460 &    \nodata &        out &  9.30$\pm$0.02 &        out &     out & 2011 Apr 19 \\
J04554822+3020165 &       HD 31305  &  6.51$\pm$0.02 &  6.38$\pm$0.02 &  6.10$\pm$0.03 &  5.18$\pm$0.03 & 2004 Feb 14 \\
                  &            &        out &  6.44$\pm$0.02 &        out &  5.21$\pm$0.03 & 2008 Nov 1 \\
J04571766+1525094 &       HD 286178   &  7.69$\pm$0.02 &  7.73$\pm$0.02 &  7.65$\pm$0.03 &  7.66$\pm$0.03 & 2004 Sep 7 \\
\enddata
\tablecomments{Entries of ``$\cdots$" and ``out"
indicate measurements that are absent because of non-detection and
a position outside the field of view of the camera, respectively.}
\tablenotetext{a}{2MASS Point Source Catalog.}
\end{deluxetable}

\clearpage
\begin{deluxetable}{llrl}
\tabletypesize{\scriptsize}
\tablewidth{0pt}
\tablecaption{New MIPS 24~\micron\ Photometry for Members of Taurus\label{tab:mips}}
\tablehead{
\colhead{2MASS\tablenotemark{a}} &
\colhead{Name} &
\colhead{[24]} &
\colhead{Date} 
}
\startdata
J04034997+2620382 &      XEST 06-006   & \nodata & \nodata \\
J04035084+2610531 &          HBC 359   &  8.98$\pm$0.11 & 2009 Mar 25 \\
J04105425+2501266 &    \nodata &  6.03$\pm$0.04 & 2004 Sep 23 \\
J04144158+2809583 &    \nodata & \nodata & \nodata \\
J04153452+2913469 &    \nodata &  3.52$\pm$0.04 & 2007 Feb 23 \\
J04153566+2847417 &    \nodata &  5.62$\pm$0.04 & 2007 Feb 23 \\
J04154131+2915078 &    \nodata & \nodata & \nodata \\
J04154269+2909558 &    \nodata & \nodata & \nodata \\
J04154807+2911331 &    \nodata & \nodata & \nodata \\
J04161726+2817128 &    \nodata &  9.16$\pm$0.10 & 2004 Sep 23 \\
                  &            &  9.29$\pm$0.15 & 2007 Feb 23 \\
J04162725+2053091 &    \nodata & \nodata & \nodata \\
J04163048+3037053 &    \nodata & \nodata & \nodata \\
          \nodata &  SST Tau 041831.2+282617  &  6.75$\pm$0.04 & 2004 Sep 23 \\
                  &            &  6.69$\pm$0.04 & 2005 Feb 28 \\
                  &            &  6.56$\pm$0.04 & 2007 Feb 23 \\
J04185813+2812234 &   IRAS 04158+2805   &  2.87$\pm$0.04 & 2008 Mar 13 \\
J04213965+2649143 &    \nodata & \nodata & \nodata \\
J04215482+2642372 &    \nodata & \nodata & \nodata \\
J04215851+1520145 &    \nodata & \nodata & \nodata \\
J04242321+2650084 &    \nodata &  7.82$\pm$0.06 & 2005 Feb 27 \\
J04245021+2641006 &    \nodata & \nodata & \nodata \\
J04251550+2829275 &    \nodata & \nodata & \nodata \\
J04264449+2756433 &    \nodata & \nodata & \nodata \\
J04270739+2215037 &    \nodata & \nodata & \nodata \\
J04272467+2624199 &    \nodata & \nodata & \nodata \\
J04314644+2506236 &    \nodata & \nodata & \nodata \\
J04324107+1809239 &  RXJ 0432.7+1809  &  6.39$\pm$0.04 & 2004 Feb 20 \\
                  &            &  6.51$\pm$0.04 & 2006 Feb 19 \\
J04325323+1735337 &  RXJ 0432.8+1735  &  6.51$\pm$0.04 & 2004 Sep 25 \\
J04340619+2418508 &    \nodata & \nodata & \nodata \\
J04345973+2807017 &    \nodata & \nodata & \nodata \\
J04355760+2253574 &    \nodata &  7.03$\pm$0.05 & 2005 Mar 1 \\
                  &            &  7.25$\pm$0.06 & 2007 Feb 26 \\
J04355881+2438404 &    \nodata &  9.38$\pm$0.24 & 2005 Mar 1 \\
J04355949+2238291 &    \nodata &  5.38$\pm$0.05 & 2005 Mar 1 \\
                  &            &  5.20$\pm$0.04 & 2007 Feb 26 \\
J04363248+2421395 &    \nodata & \nodata & \nodata \\
J04380191+2519266 &    \nodata &  8.59$\pm$0.09 & 2005 Mar 4 \\
J04383907+1546137 &       HD 285957   &  7.79$\pm$0.06 & 2005 Mar 3 \\
J04554822+3020165 &       HD 31305  &  3.34$\pm$0.04 & 2004 Sep 25 \\
                  &            &  3.36$\pm$0.04 & 2004 Oct 12 \\
                  &            &  3.33$\pm$0.04 & 2005 Mar 5 \\
J04571766+1525094 &       HD 286178   &  7.40$\pm$0.05 & 2004 Oct 13 \\
J05064662+2104296 &    \nodata & \nodata & \nodata \\
\enddata
\tablecomments{Entries of ``$\cdots$" 
indicate measurements that are absent because of non-detection.}
\tablenotetext{a}{2MASS Point Source Catalog.}
\end{deluxetable}

\clearpage
\begin{deluxetable}{c}
\tablecolumns{1}
\tabletypesize{\scriptsize}
\tablewidth{0pt}
\tablecaption{Boundaries for Figure \ref{fig:candidate} \label{tab:bound}}
\tablehead{
\colhead{(x,y)} 
}
\startdata
\cutinhead{$W1$ vs. $\mw1 - \mw2$} 
$(0.12, 6.00)$ \\ 
 $(0.79, 15.9)$ \\ 
 $(0.41, 15.9)$ \\ 
 $(2.50, 15.9)$ \\
 (-0.1, 8.50)  \\
 $(2.50, 6.00)$ \\
 (-0.1, 6.00)  \\
 \cutinhead{$\mw1-\mw3$ vs. $\mw1-\mw2$} 
  (-0.1, 1.10) \\
  (-0.1, -0.50) \\
  (-0.1, 1.85) \\
   (0.56, -0.5) \\
 $(1.45, 6.80)$\\ 
  $(0.56, 1.10)$ \\
  $(2.50, 6.80)$ \\
  $(2.50, 4.60)$   \\
  $(0.43, 1.10)$  \\
 \cutinhead{$\mw1-\mw4$ vs. $\mw1-\mw2$} 
  $(2.50, 8.80)$ \\ 
(-0.1, 0.40) \\
 $(1.45, 10.5)$ \\
 $(0.25, 0.90)$ \\
  $(2.50, 10.5)$ \\
 $(0.55, 2.90)$ \\ 
  $(0.55, 2.90)$\\
  $(0.25, 0.90)$ 
\enddata
\end{deluxetable}

\clearpage

\begin{deluxetable}{lccccccc}
\tablecolumns{7}
\tabletypesize{\scriptsize}
\tablewidth{0pt}
\tablecaption{IR Excess Sources that are Probable Non-members\label{tab:rej}}
\tablehead{
\colhead{} &
\colhead{} &
\multicolumn{3}{c}{Satisfied membership criterion?} &
\colhead{} &
\colhead{} \\
\cline{3-5} 
\colhead{WISE\tablenotemark{a}} & 
\colhead{2MASS\tablenotemark{b}} &
\colhead{USNO \&} &
\colhead{IRAC} &
\colhead{$\mu$} &
\colhead{Visual} &
\colhead{Spectra\tablenotemark{d}} \\
\colhead{} &
\colhead{} &
\colhead{2MASS} &
\colhead{} &
\colhead{} &
\colhead{Inspection\tablenotemark{c}} &
\colhead{}
}
\startdata
  J040000.62+223340.9 & J04000062+2233409 & N & \nodata & \nodata & extended & \nodata \\
  J040002.50+223245.6 & \nodata & N & \nodata & \nodata & extended & \nodata \\
  J040013.58+305903.4 & J04001359+3059035 & Y & \nodata & \nodata & \nodata & M5 \\
  J040019.05+304121.3 & J04001906+3041213 & N & \nodata & \nodata & \nodata & galaxy \\
  J040022.18+303221.1 & J04002219+3032212 & Y & \nodata & N & \nodata & M2 \\
  \enddata
\tablecomments{Table \ref{tab:rej} is available in a machine-readable format, a portion is shown here for guidance regarding its form and content.}
\tablenotetext{a}{Coordinate-based identifications from the \wise All-Sky
Source Catalog.}
\tablenotetext{b}{Coordinate-based identifications from the 2MASS Point Source
Catalog when available.}
\tablenotetext{c}{Probable non-members based on:
false = spurious \wise detection in all bands or bands that seem to show excesses;
mismatch = different sources dominate in $W1$/$W2$ and $W3$/$W4$;
galaxy = resolved galaxy;
extended = extended in 2MASS, DSS or SDSS images, indicating that it may be a galaxy;
blend = unreliable photometry because of blending with other sources.}
\tablenotetext{d}{Probably non-members based on the listed spectroscopic classification}
\end{deluxetable}
\clearpage

\begin{deluxetable}{lccc}
\tablecolumns{4}
\tabletypesize{\scriptsize}
\tablewidth{0pt}
\tablecaption{IR Excess Sources Observed with Spectroscopy\label{tab:spec}}
\tablehead{
\colhead{2MASS\tablenotemark{a}} &
\colhead{Spectral} &
\colhead{$W_\lambda$(H$\alpha$)} &
\colhead{Instrument}  \\
\colhead{} & \colhead{Type} & \colhead{(\AA)} &\colhead{}
}
\startdata
\cutinhead{New Members}
J04064443+2540182 &M5.75 & $15\pm1$&LRS \\
J04102834+2051507 & M5.5 &\nodata&SpeX\\
J04124068+2438157  & M4 &  \nodata& SpeX \\
J04215851+1520145 & M4 & \nodata& SpeX \\
J04284199+1533535 & M5 &\nodata & SpeX \\
J04285053+1844361  & M7.25 & $48\pm5$&LRS \\
J04332789+1758436 & M & \nodata& SpeX \\
J04343128+1722201\tablenotemark{b} & M4.25 & $105\pm5$&LRS \\
J04360131+1726120\tablenotemark{b} & M3 &$23.5\pm1$&LRS \\
J04374333+3056563 & M3.75 &$19\pm1$&LRS \\
J04384502+1737433\tablenotemark{b} & M4.25 & $57\pm5$&LRS \\
J04384725+1737260 & M5.5 & $42\pm2$&LRS \\
J04440164+1621324\tablenotemark{c} & M6 &\nodata&SpeX \\
J04480632+1551251 & M4.5 &\nodata& SpeX \\
J04481348+2924537 & M1.75 & $18.5\pm1$&LRS\\
J04485745+2913521\tablenotemark{d} & M6 &\nodata&SpeX\\
J04485789+2913548 & M & \nodata & SpeX \\
J04504003+1619460\tablenotemark{b} & M4.75 &$17\pm1$& LRS \\
J04520970+3037454\tablenotemark{e} &  M &  \nodata& SpeX \\
J04585141+2831241\tablenotemark{f} & M2.5+M5.5 & \nodata & SpeX \\
J04591661+2840468  & \nodata & \nodata & SpeX \\
J05023483+2745499\tablenotemark{b}  & M4.25 & $10\pm0.5$&LRS\\
J05023985+2459337\tablenotemark{b}  & M4.25 & $5\pm0.5$&LRS \\
J05073903+2311068  & M4.5 &  \nodata& SpeX \\
J05080709+2427123  & M4 &  \nodata& SpeX \\
\cutinhead{Non-members}
J04001359+3059035\tablenotemark{g} & M5 & \nodata& SpeX \\
J04064531+1759436& M4V &  \nodata& SpeX \\
J04094207+2128000\tablenotemark{h} & M1+early &  \nodata& SpeX \\
J04135394+1634015 & M3V &  \nodata& SpeX \\
J04152336+3006258& M3.5V &  \nodata& SpeX \\
J04183835+2719596 & $<$M0 & \nodata& SpeX \\
J04185139+2531380 & M3V &  \nodata& SpeX \\
J04214429+1929454 & $<$M0 & \nodata& SpeX \\
J04221376+1525298 &F2-F4 & \nodata& SpeX \\
J04222131+2037524 & M5III &   \nodata& SpeX \\
J04332503+1526460 & galaxy & \nodata & SpeX \\
J04401083+1628586 &$<$M0 &\nodata& SpeX \\
J04403220+2959143 & M5III & \nodata& SpeX \\
J04404770+2739469 & galaxy & \nodata& SpeX \\
J04420028+2042566 & M3V &  \nodata& SpeX \\
J04422696+2941426 & galaxy & \nodata & SpeX \\
J04431538+2444558 & M6III & \nodata & SpeX \\
J04482211+3018479 & early & \nodata & SpeX \\
J04494524+2001529 & galaxy &  \nodata& SpeX \\
J04503102+1514127& M2.5V &  \nodata& SpeX \\
J04531985+1915276 & galaxy & \nodata & SpeX \\
J05024059+1922373 & M5III & \nodata & SpeX \\
J05041551+2702196 & galaxy & \nodata & SpeX \\
J05043642+2716380 & galaxy & \nodata & SpeX \\
J05064796+1922501 & M5III & \nodata & SpeX \\
J05062981+2245367\tablenotemark{h}  & late K/early M + M2.5V & \nodata & SpeX \\
J05071092+2602125 & galaxy & \nodata & SpeX \\
\enddata
\tablenotetext{a}{Coordinate-based identifications from the 2MASS Point Source
Catalog.}
\tablenotetext{b}{UCAC4 proper motion measurements consistent with Taurus membership.}
\tablenotetext{c}{The PPMXL \citep{roe10} proper motions are inconsistent
with Taurus, but based on six astrometric positions (UKIDSS, WISE, 2MASS,
The Guide Star Gatalog, USNO-B1.0, APM-North; \citealt{GSC}), 
we calculate a proper motion of
$\mu_{\alpha}q, \mu_{\delta}$ = 14\,$\pm$\,4, -19\,$\pm$\,4
mas/yr, which is in good agreement with that of the neighboring L1551 cluster
\citep[Tau VI;][]{luh09}.}
\tablenotetext{d}{Companion to 2MASS J04485789+2913548}
\tablenotetext{e}{Haro 6-39}
\tablenotetext{f}{BS Tau A+B}
\tablenotetext{g}{The spectrum indicates that it is young, but it is probably too far from the known Taurus population to be a member.}
\tablenotetext{h}{Spectra obtained for both components of a close pair.}
\end{deluxetable}

\clearpage

\begin{deluxetable}{lrrrr}
\tablecolumns{5}
\tabletypesize{\scriptsize}
\tablewidth{0pt}
\tablecaption{IR Excess Sources with Undetermined Membership in Taurus\label{tab:unknown}}
\tablehead{
\colhead{WISE\tablenotemark{a}} &
\colhead{$W1$} &
\colhead{$W2$} &
\colhead{$W3$} &
\colhead{$W4$} %&
}
\startdata
  J040442.58+290211.8 & $13.71\pm 0.03$ & $12.58\pm 0.03$ & $10.47\pm 0.09$ & $8.29\pm 0.27$  \\
  J040802.82+292331.4 & $13.62\pm 0.03$ & $12.63\pm 0.03$ & $9.78\pm 0.06$ & $7.29\pm 0.12$  \\
  J041422.20+193626.8 & $13.92\pm 0.03$ & $12.61\pm 0.03$ & $9.97\pm 0.05$ & $8.12\pm 0.23$  \\
  J041858.20+193901.3 & $13.24\pm 0.03$ & $12.07\pm 0.02$ & $9.71\pm 0.05$ & $7.55\pm 0.16$  \\
  J041955.26+290613.4 & $13.81\pm 0.03$ & $12.08\pm 0.03$ & $9.17\pm 0.05$ & $6.93\pm 0.10$  \\
  J042230.61+192332.7 & $13.50\pm 0.03$ & $12.17\pm 0.02$ & $8.84\pm 0.03$ & $6.28\pm 0.05$  \\
  J042252.96+243618.5 & $13.83\pm 0.03$ & $12.62\pm 0.04$ & $9.96\pm 0.06$ & $7.79\pm 0.20$  \\
  J042408.29+172531.4 & $13.61\pm 0.03$ & $12.61\pm 0.03$ & $9.72\pm 0.04$ & $7.15\pm 0.09$  \\
  J042430.06+245254.2 & $13.93\pm 0.03$ & $12.46\pm 0.04$ & $9.26\pm 0.05$ & $7.09\pm 0.18$  \\
  J042734.86+190427.1 & $13.69\pm 0.03$ & $12.22\pm 0.03$ & $9.29\pm 0.04$ & $7.03\pm 0.12$  \\
  J042737.75+245928.4 & $13.65\pm 0.03$ & $12.56\pm 0.04$ & $9.59\pm 0.07$ & $7.08\pm 0.15$  \\
  J042849.81+232640.5 & $13.96\pm 0.03$ & $12.61\pm 0.04$ & $9.70\pm 0.08$ & $7.30\pm 0.23$  \\
  J043013.06+293615.5 & $13.90\pm 0.03$ & $12.64\pm 0.03$ & $9.74\pm 0.06$ & $7.92\pm 0.26$  \\
  J043058.15+180329.6\tablenotemark{b} & $13.74\pm 0.03$ & $12.47\pm 0.04$ & $10.31\pm 0.11$ & $7.71\pm 0.27$  \\
  J043152.94+282605.3 & $13.54\pm 0.03$ & $12.41\pm 0.03$ & $9.69\pm 0.06$ & $7.23\pm 0.12$  \\
  J043426.11+201559.9 & $14.00\pm 0.03$ & $12.84\pm 0.04$ & $10.51\pm 0.12$ & $7.52\pm 0.18$  \\
  J043755.35+163611.4 & $13.99\pm 0.03$ & $12.50\pm 0.03$ & $9.67\pm 0.07$ & $7.04\pm 0.13$  \\
  J044023.91+292218.5 & $13.50\pm 0.03$ & $12.45\pm 0.03$ & $9.53\pm 0.04$ & $7.35\pm 0.13$  \\
  J044118.52+293616.2 & $13.86\pm 0.03$ & $12.60\pm 0.03$ & $9.52\pm 0.04$ & $6.89\pm 0.08$  \\
  J044337.55+210508.7 & $13.13\pm 0.03$ & $11.42\pm 0.02$ & $8.14\pm 0.02$ & $6.16\pm 0.05$  \\
  J044508.99+250831.1 & $13.67\pm 0.03$ & $12.67\pm 0.03$ & $9.99\pm 0.06$ & $7.44\pm 0.15$  \\
  J045018.05+203636.8 & $13.44\pm 0.03$ & $12.45\pm 0.03$ & $9.57\pm 0.04$ & $7.28\pm 0.12$  \\
  J045037.11+201059.3 & $13.64\pm 0.03$ & $12.38\pm 0.03$ & $9.84\pm 0.06$ & $7.27\pm 0.12$  \\
  J045123.89+261627.3 & $13.72\pm 0.03$ & $12.39\pm 0.03$ & $9.65\pm 0.06$ & $7.49\pm 0.18$  \\
  J045301.44+255349.5 & $13.69\pm 0.03$ & $12.68\pm 0.03$ & $9.74\pm 0.05$ & $7.37\pm 0.13$  \\
  J045333.41+200249.3 & $13.95\pm 0.03$ & $12.67\pm 0.03$ & $9.56\pm 0.04$ & $7.72\pm 0.16$  \\
  J045429.42+280135.7 & $13.62\pm 0.03$ & $12.07\pm 0.02$ & $8.75\pm 0.03$ & $6.53\pm 0.07$  \\
  J045910.18+262621.5 & $13.76\pm 0.03$ & $12.62\pm 0.03$ & $9.69\pm 0.05$ & $6.93\pm 0.10$  \\
  J045945.80+263434.7 & $13.80\pm 0.03$ & $12.70\pm 0.03$ & $9.25\pm 0.04$ & $6.44\pm 0.07$  \\
  J050158.72+150819.5 & $13.28\pm 0.03$ & $11.88\pm 0.03$ & $8.99\pm 0.03$ & $6.93\pm 0.08$  \\
  J050317.83+295910.3 & $13.50\pm 0.03$ & $12.14\pm 0.02$ & $9.36\pm 0.04$ & $6.86\pm 0.09$  \\
  J050343.07+263004.0 & $13.94\pm 0.03$ & $12.61\pm 0.03$ & $9.41\pm 0.04$ & $7.09\pm 0.10$  \\
  J050619.25+292717.5 & $13.88\pm 0.03$ & $12.63\pm 0.03$ & $9.93\pm 0.06$ & $7.31\pm 0.11$  \\
  J050902.08+302841.3 & $13.95\pm 0.03$ & $12.50\pm 0.03$ & $10.20\pm 0.07$ & \nodata  \\
  \enddata
\tablenotetext{a}{Coordinate-based identifications from the \wise All-Sky
Source Catalog.}
\tablenotetext{b}{Classified as a young stellar object by \cite{gut09}.}
\end{deluxetable}

\clearpage

\begin{deluxetable}{c}
\tablecolumns{1}
\tabletypesize{\scriptsize}
\tablewidth{0pt}
\tablecaption{Boundaries for Figure \ref{fig:usno} \label{tab:ubound}}
\tablehead{
\colhead{(x,y)} 
}
\startdata
\cutinhead{$K_s$ vs. $B - K_s$}  $(3.00, 7.50)$\\
 $(3.50, 11.0)$ \\
 $(9.00, 15.0)$ \\
\cutinhead{$K_s$ vs. $R - K_s$} $(1.25, 7.50)$\\
 $(2.50, 11.5)$ \\
 $(6.50, 15.0)$ \\
\cutinhead{$K_s$ vs. $I - K_s$} $(0.00, 7.50)$ \\
 $(2.00, 7.50)$ \\
 $(6.00, 15.0)$ \\
\cutinhead{$W1$ vs. $B- \mw1$}  $(3.00, 7.00)$ \\
 $(3.50, 11.0)$ \\
 $(8.50, 13.5)$ \\
\cutinhead{$W1$ vs. $R-\mw1$}  $(1.00, 7.00)$ \\
 $(2.50, 11.0)$ \\
 $(4.50, 14.0)$ \\
\cutinhead{$W1$ vs. $I-\mw1$}  $(0.00,7.00)$ \\
 $(2.00, 12.0)$ \\
 $(5.50, 14.0)$
\enddata
\end{deluxetable}

\clearpage

\end{document}